\documentclass[12pt]{article}
\pdfoutput=1 
\usepackage{jheppub}
\usepackage{amsmath}
\usepackage{amssymb}
\usepackage{graphicx}
\usepackage{mathrsfs}
\usepackage[utf8]{inputenc}

\usepackage{subcaption}
\usepackage{float}
\usepackage{afterpage}
\newcount\xrefpos \xrefpos=0
\newcount\yrefpos \yrefpos=0
\newcount\xput \xput=0
\newcount\yput \yput=0

\def\refpos#1 #2 #3{\global\xrefpos=#1 \global\yrefpos=#2
                         \rlap{$\smash{#3}$}}
\def\put #1 #2 #3{\xput=#1 \yput=#2
                  \advance\xput by -\xrefpos
                  \advance\yput by -\yrefpos
                  \rlap{\kern\the\xput truebp
                        \vbox to 0pt{\vss\hbox{$\displaystyle #3$}
                        \kern\the\yput truebp}}}
\def\beginlabels\refpos#1\endlabels{\hbox{$\refpos#1$}}

\newcommand{\be}{\begin{equation}}
\newcommand{\ee}{\end{equation}}
\newcommand{\bea}{\begin{eqnarray}}
\newcommand{\eea}{\end{eqnarray}}
\def\bse{\begin{subequations}}
\def\ese{\end{subequations}}

\def\IZ{\relax\ifmmode\hbox{Z\kern-.4em Z}\else{Z\kern-.4em Z}\fi}

\def\AdS#1{AdS$_{#1}$}

\definecolor{rust}{rgb}{0.8,0.2,0.2}

\title{Dynamics of Holographic Entanglement Entropy Following a Local Quench}

\author[a]{Mukund Rangamani}
\author[b]{\!, Moshe Rozali}
\author[b]{\!, Alexandre Vincart-Emard}
\affiliation[\,a]{
Center for Quantum Mathematics and Physics (QMAP)  \\
Department of Physics, University of California, Davis, CA 95616 USA}
\affiliation[\,b]{
Department of Physics and Astronomy, University of British Columbia,\\
Vancover, BC V6T 1Z1, Canada}
\emailAdd{mukund@physics.ucdavis.edu}
\emailAdd{rozali@phas.ubc.ca}
\emailAdd{ave@phas.ubc.ca}
\abstract{
We discuss the behaviour of holographic entanglement entropy following a local quench in  2+1 dimensional strongly coupled CFTs. The entanglement generated by the quench propagates along an emergent light-cone, reminiscent of the Lieb-Robinson light-cone propagation of correlations in non-relativistic systems.  We find the speed of propagation is bounded from below by the entanglement tsunami velocity obtained earlier for global quenches in holographic systems, and from above by the speed of light. The former is realized for sufficiently broad quenches, while the latter pertains  for well localized quenches. The non-universal behavior in the intermediate regime appears to stem from finite-size effects. We also note that the entanglement entropy of subsystems  reverts to the equilibrium value exponentially fast, in contrast to a much slower equilibration seen in certain spin models.}

\begin{document}
\maketitle

\section{Introduction}
\label{sec:intro}
%

In recent years we have seen enormous progress in qualitative and quantitative understanding of out-of-equilibrium quantum dynamics. Theoretical and numerical methods have been very effective to unearth the generic behaviour of a variety of observables in such systems. Coupled with the rapid growth of experimental techniques in cold atom and many-body systems to probe such dynamics,  one can furthermore ratify our theoretical understanding. Motivated by these considerations we continue our explorations of dynamics of strongly coupled non-equilibrium quantum systems using holographic methods.

One simple scenario of interest in many circumstances is a situation where we start with a QFT in global  equilibrium and deform it by turning on external sources for relevant operators. The sources provide external dials which can serve to do work on the system and drive it out of equilibrium. We could consider sources that act homogeneously in space (but localized in time), which is often referred to as {\em global quench}, or have it act locally in spacetime, which corresponds to a {\em local quench}. Both types of protocols are well studied in literature in the past decade or thereabouts. In either case we are considering 
deformations of the form
\begin{align}
S_\text{QFT} \mapsto S_\text{QFT} + \int d^dx\, {\cal J}(x)\, {\cal O}(x) \,,
\label{eq:quench}
\end{align}
where ${\cal O}(x)$ is a (composite) operator of the QFT and ${\cal J}$ the classical source we dial. The distinction at this level between local and global quenches is simply in the spacetime support of the source ${\cal J}(x)$.

Much of the analytic progress in this front has been in 1+1 dimensional CFTs, where the quench protocols  of the form  \eqref{eq:quench} can be incorporated into a Euclidean path integral, and studied efficiently by computing correlation functions of the deforming operator ${\cal O}(x)$ in the unperturbed state of the CFT, cf., \cite{Calabrese:2005in,Calabrese:2006rx} for the original discussion and 
\cite{Calabrese:2007rg} for a review. 

Our primary interest is in exploring the dynamics of strongly coupled QFTs subject to such protocols in higher dimensions. A natural framework to explore this question is provided by the holographic AdS/CFT duality which maps the QFT problem onto the dynamics of a gravitational system in asymptotically AdS spacetime. For concreteness we will focus on 2+1 CFTs which are originally in global thermal equilibrium and subject them to a quench by a local scalar operator ${\cal O}$ of dimension $\Delta$. The gravitational problem then comprises of Einstein-Hilbert gravity coupled to a massive scalar, whose mass $m$ is related to the conformal dimension by the standard formula, viz.,  $\Delta = \frac{3}{2} + \sqrt{\frac{9}{4} + m^2\, \ell_\text{AdS}^2}$.\footnote{ We will only consider deformations by operators which are well separated from the unitarity bound -- our focus will be on conformally coupled scalars with $\Delta =2$.} The initial global equilibrium state maps onto a planar Schwarzschild-\AdS{4} black hole and the problem at hand involves analyzing the deformation of this said black hole consequent to turning on a boundary source for the scalar field. This then amounts to a gravitational infall problem. The pulse of scalar on the boundary propagates into the bulk and dissipates through the black hole horizon. Of interest to us are the observables in the interim process.

While there are many quantities that could be, and indeed have been \cite{Hubeny:2007xt,Chesler:2008hg,Bhattacharyya:2009uu,Das:2010yw,Hubeny:2010ry,AbajoArrastia:2010yt,Albash:2010mv,Balasubramanian:2010ce,Balasubramanian:2011ur,Aparicio:2011zy,Basu:2011ft,Balasubramanian:2011at,Keranen:2011xs,Galante:2012pv,Caceres:2012em,Buchel:2012gw,Bhaseen:2012gg,Basu:2012gg,Hubeny:2013hz,Nozaki:2013wia,Buchel:2013lla,Hartman:2013qma,Basu:2013soa,Buchel:2013gba,Liu:2013iza,Balasubramanian:2013oga,Liu:2013qca,Asplund:2013zba,Abajo-Arrastia:2014fma,Astaneh:2014fga,Buchel:2014gta,Das:2014hqa,Bai:2014tla,Rangamani:2015sha,Leichenauer:2015xra,Das:2015jka,Ecker:2015kna,Ziogas:2015aja,Sohrabi:2015qda,Camilo:2015wea}, studied in this context, we will for definiteness focus our attention on entanglement entropy. While strictly not an observable, the entanglement entropy for a particularly chosen spatial region of the QFT captures  important aspects of the field theory dynamics. Not only does it provide a measure of how correlations in the system evolve following the quench, but it furthermore is also a simple quantity to compute in the holographic context. The holographic entanglement entropy proposals of \cite{Ryu:2006bv,Ryu:2006ef} and their covariant generalization \cite{Hubeny:2007xt} provide an extremely simple route to its computation. All we are required to do is solve  a classical problem of finding areas of extremal surfaces anchored on the said region of interest. 

In what follows we will explore how holographic entanglement entropy evolves following a local quench. We will restrict our attention to a very specific scenario, wherein we quench a CFT$_3$ with a $\Delta =2$ operator. The disturbance will be taken to be localized in space  and time -- we pick exponential damping in space and an inverse P\"oschl-Teller switch on/off in time, cf., \eqref{eq:source}. We retain translational invariance in one spatial direction, breaking homogeneity in the other. We study entanglement entropy for strip-like spatial regions that are aligned with the symmetry we retain, so that the problem of finding extremal surfaces can be mapped to effectively finding geodesics in an auxiliary three dimensional spacetime. Of interest to us are how the entanglement entropy growth is correlated with the position and size of the strip relative to the quench location.

To appreciate the question, let us recall some well known facts. The classic analysis of \cite{Calabrese:2005in} of entanglement entropy growth following a global quench in CFT$_2$ has spurred lots of activity on the subject. While the two dimensional case can effectively be described by a quasiparticle picture, since the entanglement growth is linear due to left and right movers decoupling (following an initial quadratic ramp up \cite{Hubeny:2013hz,Liu:2013iza,Liu:2013qca}), the holographic models present a much different picture in higher dimensions.\footnote{ We note here that oftentimes global quenches are holographically modeled by considering a Vaidya-AdS geometry (see \cite{Danielsson:1999fa,Hubeny:2007xt} for early discussions) that corresponds to infalling null matter in the bulk, which does not accord a clean CFT interpretation. A more cleaner perspective is offered by  either solving the non-linear dynamics of gravity coupled to realistic matter like a scalar field, or more simply by implementing an end of the world brane boundary state \cite{Hartman:2013qma} explicitly in holography. The results for the growth of entanglement entropy are however independent of the particularities of the modeling.}  

The results of various analyses of global quenches have been beautifully encapsulated in the `entanglement tsunami' picture developed by Liu-Suh in \cite{Liu:2013iza,Liu:2013qca} and further explored recently in \cite{Casini:2015zua}.  Following an initial quadratic growth in time, the entanglement entropy for any region grows linearly at a rate dictated by the tsunami velocity $v_E$. To define this quantity unambiguously the authors chose to normalize the local value of entanglement entropy relative to the final thermal entropy expected for the same region once equilibration is complete. This does leave a single parameter which is the aforementioned velocity. It was found not only $v_E \leq 1$ as required by causality with equality in $d=2$ consistent with the CFT$_2$ analysis, but one could further bound it by a universal dimensional dependent constant $v_E^{*}(d)$.\footnote{ This statement as far as we are aware is robust for QFTs whose holographic duals are given in terms of two derivative Einstein-Hilbert gravity coupled to sensible matter. There is a-priori no reason for them to hold when the gravitational dynamics includes higher derivative corrections and we in particular are not aware of any statement of this kind.} This upper bound on velocity was attained holographically for matter that collapsed into a Schwarzschild-\AdS{d} black hole at late times. 

Given this rather clear situation for global quenches, we are interested in ascertaining the behaviour when we localize the quench protocol to a finite spatial domain. We in principle could focus on deformations by sources delta-function supported at point. This is natural when studying this problem in QFT as one can map the computation to that of computing correlation functions on some background, however for our purposes of carrying out numerical investigations we choose to smear out the source. We expect firstly that the underlying locality of the QFT forces entanglement entropy to behave causally; as explained in \cite{Casini:2003ix,Headrick:2014cta} this means that the source makes its presence felt only when it acts in the causal past of the entangling surface (the boundary of the region of interest). This is indeed what one sees in explicit computations in  CFT$_2$. The entanglement entropy only starts changing after a time lag set by the time it takes for the quench disturbance to propagate between the region of interest and its complement. As long as the quench front is localized either in the region or in the complement, we only have the initial state entanglement.

Previous analyses of holographic local quenches by \cite{Nozaki:2013wia} involved modeling the system by the infall of a massive particle -- this is effectively an eikonal approximation wherein one is assuming that the wavepackets of the quench are tightly collimated.  Moreover, the authors chose to work with very heavy operators $\Delta \gg 1$ which could then be approximated in terms of worldlines of a small black holes. The relevant geometry can be obtained by applying a suitable symmetry transformation to the global Schwarzschild-AdS black hole and with it in hand properties of  holographic entanglement entropy were explored. This picture was further supported by field theory analysis of such deformations at large central charge \cite{Asplund:2014coa,Asplund:2015eha}. Our aim to tackle this problem from a different perspective by studying the entanglement evolution in a quenched gravitational background as explained above. We will recover most of the results mentioned above in our analysis. 

We can moreover explore quantitative features of the entanglement evolution. We see that the propagation of entanglement is confined to a an effective light-cone. We extract an entanglement velocity $v_E$ from this emergent causal structure.  Unlike the case of the global quench, the velocity depends on the details of the quench.  It appears to grow monotonically with increase in the amplitude of the quench source as well as with the increase of the initial temperature. For a certain range of parameters is appears to track the tsunami velocity bound $v_E^*(3)$ of \cite{Liu:2013iza}, while for others it reaches close to the speed of light. 

One can understand this behaviour qualitatively as follows: for well localized quenches in large regions $\mathcal{A}$ one is in the eikonal limit. Here the growth of the entanglement entropy is linear with an `entanglement velocity' that is close to the speed of light. On the other hand for regions which are confined within a broad quenching pulse, the situation is nominally similar to a global quench experiment. As we are collapsing scalar matter, we should expect that the behaviour in this domain is isomorphic that seen by \cite{Liu:2013iza}, and indeed we recover the tsunami velocity.\footnote{ There is a somewhat annoying fact that the tsunami velocity $v^*_E(3) = 0.687$ in three spacetime dimensions is  mraginally lower than the speed of sound $v_s = 0.707 $, making it somewhat hard to convincingly  point to precise origin of the effect. We nevertheless feel confident that the analogy with the global quench points to the tsunami velocity being the operative feature.} Away from these limiting cases we see contamination from edge effects both from finite size of $\mathcal{A}$ and the finite width of the quench source. We have not examined the detailed non-linear effects that cause the velocity to grow from the tsunami bound towards the speed of light, but display some examples which illustrate the pattern.

While our numerical results are constrained to probing small spatial regions relative to thermal scale,\footnote{ This constraint arises because our numerical solutions only determine the geometry to the exterior of the apparent horizon. For small regions ${\cal A}$ the extremal surfaces stay in this domain, but for larger regions, they do penetrate the apparent horizon -- see \cite{AbajoArrastia:2010yt,Hubeny:2013dea}.} we nevertheless are able to extract both this entanglement velocity as well as examine the return to the equilibrium. In contrast to studies in lattice models in low dimensions which display a logarithmic return of entanglement entropy to its equilibrium value after the quench, we find that the holographic systems prefer to equilibrate exponentially.

The outline of the paper is as follows. In \S\ref{sec:setupquench} we describe the basic set-up for holographic local quenches, describing the general methodology and the determination of entanglement entropy from the gravitational background. In \S\ref{sec:geometry} we give the basic numerical results for the quench spacetime and extremal surfaces therein. The key statements regarding the behaviour of entanglement entropy in a locally quenched CFT are then extracted in \S\ref{sec:results}, where we describe the growth velocity $v_E$ and the return to equilibrium. We end with some open questions in \S\ref{sec:discuss}.  Some details of the numerical methods are collected in the Appendices.

\section{Preliminaries: Holographic Local Quench}
\label{sec:setupquench}

We are interested in the behaviour of entanglement entropy in a $2+1$ dimensional field theory that has been driven out of equilibrium locally by an inhomogeneous relevant scalar operator. Holographically, this amounts to solving the gravitational dynamics of a $3+1$ dimensional asymptotically AdS spacetime and its consequences for the area of extremal surfaces anchored on the boundary.

\subsection{Metric Ansatz}
\label{sec:ansatz}

In order to dynamically evolve a spacetime geometry following a local quench, it is convenient to choose our metric ansatz to be a generalization of the infalling Eddington-Finkelstein coordinates for black holes. We choose to work in an asymptotically \AdS{4} spacetime, dual to a $2+1$ dimensional CFT,
\begin{align}
 ds^2 = - 2 \,A\, e^{2 \chi} \, dt^2 + 2 \, e^{2 \chi }\, dt \, dr - 2\, F_x \,  dt dx + \Sigma^2 \left( e^{B} \, dx^2 + e^{-B} \,dy^2 \right), 
 \label{ansatz1}
\end{align}
where $r$ denotes the radial bulk coordinate, with the boundary lying at $r = \infty$, and $t$ is a null coordinate that coincides with time on the boundary. We have chosen our quench to be localized in the $x$-direction and translationally invariant in the $y$ direction. Hence all the fields appearing above $\{A, \chi, F_x, \Sigma, B\}$ depend only on the coordinates $\{r, t, x\}$ with $\partial_y$ being an isometry.

This choice for the metric has many advantages: it provides us with coordinates that remain regular throughout the entire domain as the spacetime equilibrates, it leads to a characteristic formulation of our gravitational infall problem, and it comes with a residual radial diffeomorphism that is of great computational help \cite{Chesler:2013lia}.  Indeed, the metric \eqref{ansatz1}  remains invariant under radial shifts,\footnote{ For notational clarity, we use upper case Latin indices $\{M, N, ...\}$ to represent bulk coordinates, and lower case Greek indices $\{ \mu, \nu, ...\}$ to refer to boundary coordinates.}
\begin{align}
r \rightarrow \overline{r} = r + \lambda(x^\mu) \,.
\label{eq:rrepar}
\end{align}
On physical grounds, we anticipate that the black hole's horizon will grow locally as the effects of matter from the boundary are felt in the interior of the bulk. Hence a sensible gauge choice is to dynamically determine $\lambda$ so that the coordinate location of the black hole's apparent horizon\footnote{ See Appendix A for further details about our numerical scheme} remains fixed. This keeps the calculational domain simple.

Einstein's equations in the presence of a scalar field are given by 
\begin{equation}
\begin{split}
& R_{MN} - \frac{R}{2} G_{MN} - \frac{d(d-1)}{2\,\ell_\text{AdS}^2} G_{MN} = T_{MN}
 \\
& T_{MN} = \nabla_M \Phi \nabla_N \Phi + G_{MN} \mathcal{L}_\Phi, \;\;\;\;\; \mathcal{L}_\Phi = -\frac{1}{2} \left( G^{MN} \nabla_M \Phi \nabla_N \Phi + m^2 \Phi^2 \right) .
\end{split}
\end{equation}
For simplicity, we restrict our attention to $m^2 \,\ell_\text{AdS}^2= -2$ so that the asymptotic expansion of the scalar field near the boundary is analytic in powers of $1/r$:
\begin{align}
\Phi(r,t,x) = \frac{\phi_0(t,x)}{r} + \frac{\phi_1(t,x)}{r^2} + \cdots
\end{align}
We note that since $t$ is a null coordinate, $\phi_1(t,x)$ will have contributions coming from both the source and the response of the scalar field, as will be explained below.

\subsection{Asymptotic Geometry}
\label{sec:asympt}

In a theory of gravity on asymptotically AdS spacetimes, asymptotic analysis alone is not sufficient to determine the bulk metric \cite{deHaro:2000xn}. Indeed, the missing piece in the asymptotic analysis is the boundary stress tensor, determined by solving the full bulk equations:
\begin{align}
T_{\mu \nu} \sim g_{\mu \nu}^{(3)} \,,
\end{align}
where $g_{\mu \nu}^{(3)}$ is the part of the metric undetermined by the equations of motion for $d=3$.

While our infalling coordinate chart \eqref{ansatz1} differs from the standard Fefferman-Graham chart typically used for asymptotic expansions, it is a straightforward exercise to carry out an asymptotic analysis. Demanding that the field equations are obeyed in the near-boundary $r \to \infty$ domain we find
\begin{align}  
A(r,t,x) =& \;\; \frac{(r+\lambda(t,x))^2}{2} - \partial_t \lambda(t,x) - \frac{1}{4} \phi_0(t,x)^2+ \frac{a^{(3)}(t,x)}{r} + \cdots \\
                     \chi(r,t,x) =& \;\;\frac{c^{(3)}(t,x)}{r^3} + \cdots \\
                     F_x(r,t,x) =& \;\;- \partial_x \lambda(t,x) + \frac{f^{(3)}(t,x)}{r} + \cdots \\
                     \Sigma(r,t,x) =& \;\;r+\lambda(t,x) - \frac{1}{4} \phi_0(t,x)^2 +  \cdots \\
                     B(r,t,x) =& \;\;\frac{b^{(3)}(t,x)}{r^3} + \cdots \,.
\end{align}
One may also show that the explicit map to the Fefferman-Graham coordinates $\{\tau,\rho,\xi\}$ takes the asymptotic form 
\begin{align}  
\tau(r,t,x) =& \;\; t + \frac{1}{r} - \frac{\lambda(t,x)}{r^2} + \cdots, \\
\rho(r,t,x) =& \;\; r + \lambda(t,x) - \frac{1}{4} \frac{\phi_0(t,x)^2}{r} + \cdots, \\
\xi(r,t,x) =& \;\; x + \mathcal{O}(r^{-3}) . 
\end{align}
Additional care needs to be taken when dealing with scalar fields in a theory of gravity formulated in terms of null coordinates. Indeed, the falloff of scalar fields with $m^2 \ell_\text{AdS}^2 = -2$ is known to behave in Fefferman-Graham coordinates as: 
\begin{align} 
\Phi(\rho, \tau, \xi) = \frac{\phi_\text{source}(\tau,\xi)}{\rho} + \frac{\phi_\text{response}(\tau,\xi)}{\rho^2} + \cdots 
\end{align}
as we approach $\rho \rightarrow \infty$. By using the coordinate expansion above, we obtain%
\begin{align} 
\Phi(r,t,x) = \frac{\phi_\text{source}(t,x)}{r} + \frac{\phi_\text{response}(t,x)  + \partial_t \phi_\text{source}(t,x) - \lambda(t,x) \phi_\text{source}(t,x) }{r^2} + \cdots, 
\end{align}
thus confirming our earlier claim that $\phi_1 = \phi_\text{response}  + \partial_t \phi_0 - \lambda \,\phi_0 $ mixes the source and the expectation value of the scalar. 

\subsection{Boundary Stress Tensor}
\label{sec:Tmn}

In order to solve Einstein's equations as efficiently as possible, we found it useful to use the boundary stress tensor and its conservation equations to find and propagate the undetermined fields $a^{(3)}$ and $f^{(3)}$ accurately in time (in our scheme, $ b^{(3)}$ and $ c^{(3)}$ need to be read off from the solutions directly). For asymptotically \AdS{4} spacetimes, the boundary stress tensor in the presence of a scalar field of mass squared $m^2 \ell_\text{AdS}^2 = -2$ can be expressed in the Brown-York form as  
\begin{align}
T_{\mu \nu} = K_{\mu \nu} - K \gamma_{\mu \nu} + 2\, \gamma_{\mu \nu} - \left( {}^\gamma R_{\mu \nu} - \frac{1}{2} \,{}^\gamma R\, \gamma_{\mu \nu} \right) + \frac{1}{2} \, \gamma_{\mu \nu} \, \phi^2,
\end{align}
where we have introduced some boundary data:  $\gamma_{\mu\nu}$ is the induced metric on the boundary, 
$K_{\mu \nu}, K \equiv \gamma^{\mu\nu} K_{\mu\nu}$ its extrinsic curvatures, 
and ${}^\gamma R_{\mu\nu}, {}^\gamma R $ its intrinsic curvatures. Explicitly in terms of the asymptotic expansion coefficients we find that the energy-momentum tensor takes the form 
\begin{align} T_{00} =& \;\; 2 a^{(3)} + 4 c^{(3)} + \phi_0 \phi_\text{response}, \\
                    T_{tx} =& \;\; \frac{3}{2} f^{(3)} - \frac{1}{2} \phi_0 \partial_x \phi_0  \,,
\end{align}
while the conservation equations in the presence of the scalar source $\phi_0(x,t)$ read
\begin{align} 
\partial_t T_{00} =& \;\; \partial_x T_{tx} + \partial_t \phi_0 \; \phi_\text{response}, \\
\partial_t T_{tx} =& \;\; \frac{1}{2} \left( \partial_x T_{00} - 3 \; \partial_x b^{(3)} + \partial_x \phi_0  \; \phi_\text{response} - \phi_0 \; \partial_x \phi_\text{response} \right). 
\end{align}

We take our initial state  to be in thermal equilibrium, which translates to an initial condition on the bulk metric, which is then the  planar static Schwarzschild-\AdS{4}  black hole  spacetime with temperature 
\begin{align}
T = \frac{3 \, M^\frac{1}{3}}{4 \pi } \,.
\end{align}
The initial boundary stress tensor is then simply $T^\mu_{\ \nu} = \text{diag}\{1, \frac{1}{2}, \frac{1}{2}\} $.
To model our local quench, we simply need to specify a source function $\phi_0(t,x)$ and let the system evolve according to the Einstein equations, all while making sure that $\lambda$ is gauge-chosen to fix the location of the apparent horizon.

\subsection{Holographic Entanglement Entropy}
\label{sec:setupextremal}

Once we have obtained solutions for the local quench, we can study the subsequent dynamics of the entanglement entropy of a region $\mathcal{A}$ on the boundary using the covariant holographic entanglement entropy prescription \cite{Hubeny:2007xt}. The latter requires us to determine extremal surfaces anchored on the entangling surface on the boundary.

 For simplicity, we exploit the translational invariance, and restrict our attention to a strip-region
\begin{align}
\mathcal{A} = \{(x,y) \,\vert\; x \in (-L,L) , \;\; y \in \mathbb{R} \} \,,\qquad 
\partial\mathcal{A} =\{(x,y) \,\vert\; x = \pm L , \;\; y \in \mathbb{R} \} \,.
\end{align}
The extremal surfaces $\mathcal{E}_\mathcal{A}$ anchored on $\partial\mathcal{A}$ are straightforwardly determined by solving a set of ODEs. Using coordinates adapted to the $\partial_y$ isometry, we  parameterize the surface by  coordinates $y,\tau$.  Consequentially, $\mathcal{E}_{\mathcal{A}}$ is then obtained by solving the geodesic equations in an auxiliary three dimensional spacetime with metric  $\tilde{g}_{MN} dX^N\, dX^M= 
g_{yy}\, g_{MN} \, dX^N\, dX^M$,  with the restriction to $y=\text{constant}$ understood, i.e., $X^M(\tau) =\{t(\tau), r(\tau), x(\tau)\}$. Equivalently we solve the Euler-Lagrange equations obtained from the Lagrangian  $ \mathcal{L} = g_{yy} \, g_{MN} \dot{X}^M \dot{X}^N$.

While we have phrased the determination of $\mathcal{E}_\mathcal{A}$ as a boundary value problem, it is practical to switch to an initial value formulation. We parameterize the solutions by  specifying the turning point, or tip, of the geodesic in the bulk, $X^M_*(\tau) = \{t_*,r_*,x=0\}$, and evolve towards the boundary using an ODE solver (for instance the Matlab solver \textit{ode45}) until both $\partial A$ and a specified UV cutoff are reached. 

To this end, we have chosen to transform our system of 3 second order ODEs into a system of 6 first order ODEs in the variables
\begin{align} 
\left\{ t, \;\; P_t \equiv \Sigma^2 \, \dot{t}, \;\; r, \;\; P_+ \equiv e^{2 \chi}\left( \dot{r} - A \;\dot{t} \right), \;\; x, \;\; P_x \equiv \Sigma^2 \,\dot{x} - e^{-B} F_x \; \dot{t} \right\}. 
\end{align}
With these new variables,\footnote{ These definitions for the momenta ensure that all quantities are of order $\mathcal{O}(1)$ for numerical stability.} $\mathcal{L} = 2 P_+ P_t + P_x^2$. The boundary conditions at the turning point are
\begin{align} 
\left\{ t = t^*, \;\; P_t =0 , \;\; r = r^*, \;\; P_+ = 0, \;\; x = 0, \;\; P_x = \pm 1 \right\}. 
\end{align}
The conditions on $P_t$ and $P_+$ are a consequence that, because of symmetry, we expect $\dot{t} = \dot{r} = 0$ at $X^*$, whereas the condition for $P_x$ has been chosen to normalize the action by setting $\mathcal{L} = 1$. The sign determines whether the geodesic will go towards the positive or negative $x$-axis.

To translate from the length of the geodesic to the actual entanglement entropy $S_{\mathcal{A}}$ we pick an IR regulator $L_y$ along the translationally invariant direction and a UV cutoff $\epsilon$. We choose to present the results for the regulated entanglement entropy by subtracting off the corresponding answer in the unperturbed theory. There are two natural regularizations we can use:

\noindent
\underline{Regulator 1:} We subtract the entanglement in the `instantaneous thermal state' obtained by taking the Schwarzschild-\AdS{4} metric with a horizon located at $r_+(x,t) = M^\frac{1}{3} + \lambda(t,x)$.  This choice allows clean matching of the asymptotic coordinate chart.

\noindent
\underline{Regulator 2:} We alternately can choose to subtract of the vacuum entanglement entropy for the same region, with a dynamical UV cut-off $\epsilon_\text{vac}(x,t)$. This gives
\begin{align}
\Delta S_\mathcal{A} = L_y \left[\int d\tau - \frac{2}{\epsilon} - 2\,\lambda(t,x)   + \frac{4 \pi}{L} \left(\frac{\Gamma(\frac{3}{4})}{\Gamma(\frac{1}{4})} \right)^2 \right] .
\end{align}
The two regulators differ by a finite amount that is invariant temporally, allowing us to cross-check our numerical results. In what follows we will simply quote $\Delta S_\mathcal{A}$ normalized by $L_y$.

\section{The Quench Spacetime and Extremal Surfaces}
\label{sec:geometry}
We now turn to describing the results of solving Einstein's equations sources by the scalar field boundary condition. We then describe properties of the extremal surfaces of interest in these geometries. 

\subsection{Numerical Solutions}

We use the characteristic formulation of Einstein's equations resulting from the null slicing of spacetime outlined in 
\cite{Chesler:2013lia} to numerically find the geometry. Even though we start with a complicated set of PDEs, the characteristic formulation simplifies the equations of motion into two categories: the equations for the auxiliary fields that are local in time and reduce to a nested set of radial ODEs, and the equations for dynamical quantities that encode the evolution of the geometry. 

To numerically integrate the Einstein and Klein-Gordon equations, we discretize the radial direction using a Chebyshev collocation grid. This choice of discretization for the extra dimension is particularly well suited to find smooth solutions to boundary value problems while ensuring their exponential convergence as the grid size is increased. We opted to choose a rational Chebyshev basis to deal with the non-compact spatial direction. The main advantage of working with a rational Chebyshev grid is that the boundary conditions at $x= \pm \infty$ are already implemented \textit{behaviourally}; as long as the solution decays at least algebraically fast or asymptotes to a constant, we can avoid specifying the boundary conditions explicitly \cite{Boyd:2001aa}. We use a grid of 41 points in both directions. To propagate in time, we use an explicit fifth-order Runge-Kutta-Fehlberg method with adaptive step size. We also avoid aliasing in both the radial and spatial directions by applying a low-pass filter at each time step that gets rid of the top third of the Fourier modes.

We chose the source function to be $\phi_0(t,x) =  f(x) g(t)$ with
\begin{align}
f(x) = \frac{\alpha}{2} \left[ \tanh \left( \frac{x + \sigma}{4 s} \right) - \tanh \left( \frac{x - \sigma}{4 s} \right) \right], \;\;\;\;\; g(t) = \text{sech}^2 \left( \frac{t - t_q \Delta}{t_q} \right). 
\label{eq:source}
\end{align}
With it, we can ramp up the scalar field to reach its maximum value $\alpha$ at time $t = t_q  \Delta$ before it vanishes again. The parameters $\{s, t_q, \Delta \}$ are chosen to facilitate the numerics, whereas $\sigma$ determines the width of the perturbation. In practice, we found $s = 0.15$, $t_q = 0.25$ and $\Delta = 8$ to give us satisfying accuracy for the late-time behaviour of the scalar field while preserving a nicely localized shape for the pulse.
So we therefore study the quench protocols parametrized by two  parameters: an amplitude $\alpha$ and a width $\sigma$. Along with the initial temperature of the system which we take to be parametrized by $M$, we have three parameters at our disposal.
\begin{equation}
\begin{split}
\phi_0(x,t) &=  \frac{\alpha}{2} \left[ \tanh \left( \frac{5}{3}\,(x + \sigma) \right) - \tanh \left( \frac{5}{3} \, ( x - \sigma)\right) \right]  \text{sech}^2 \left( 4\, t - 8 \right)\,, 
 \\
& \qquad \text{Protocol parameters:} \; \{\alpha, \sigma, M\}
\end{split}
\label{}
\end{equation}	

\begin{figure*}[ht]
\begin{subfigure}{0.49\textwidth}
 \includegraphics[width=7.7cm]{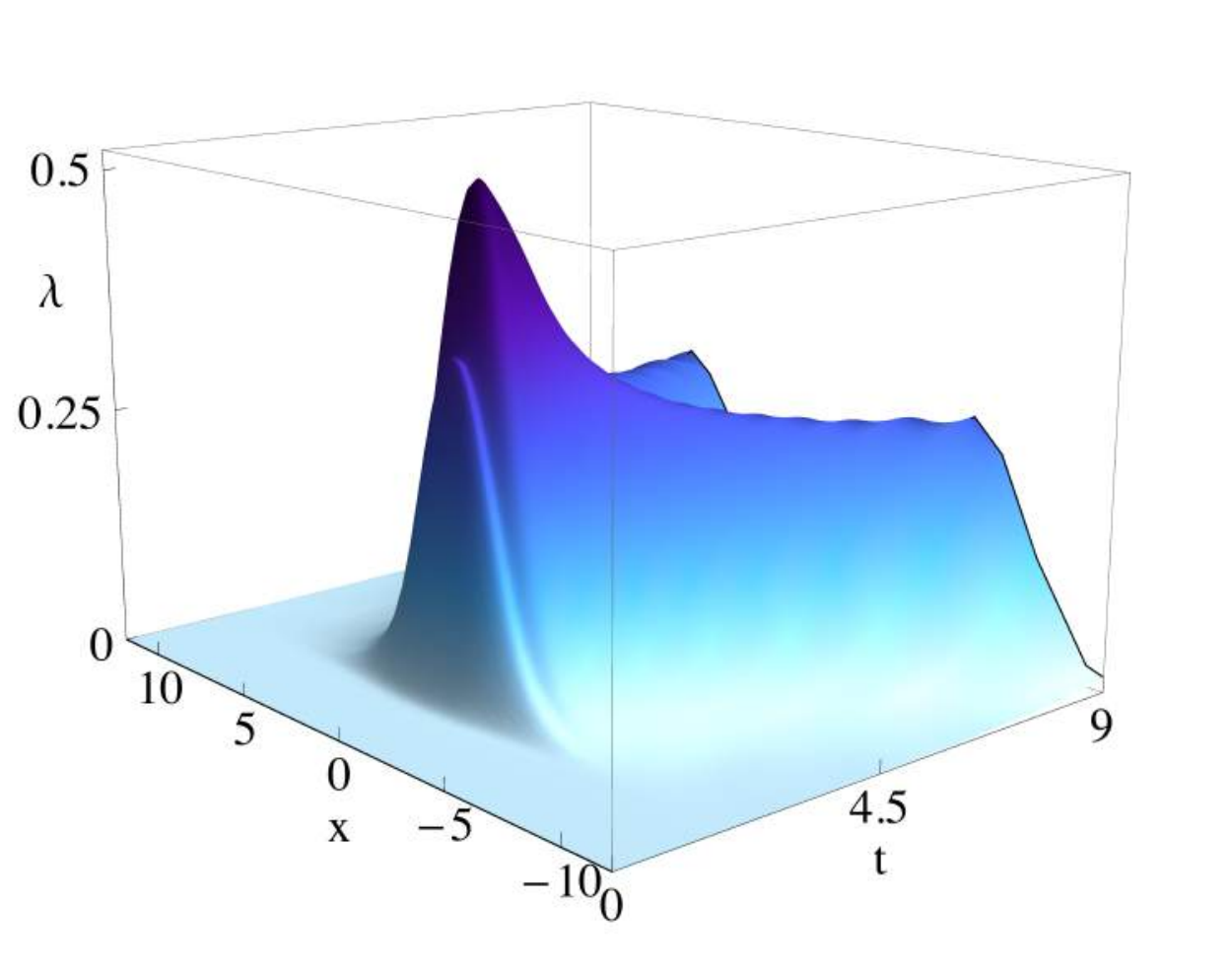}
 \subcaption{$\lambda(t,x)$}
\label{fig_entropy}
\end{subfigure}
\hfill
\begin{subfigure}{0.49\textwidth}
    \includegraphics[width=8cm]{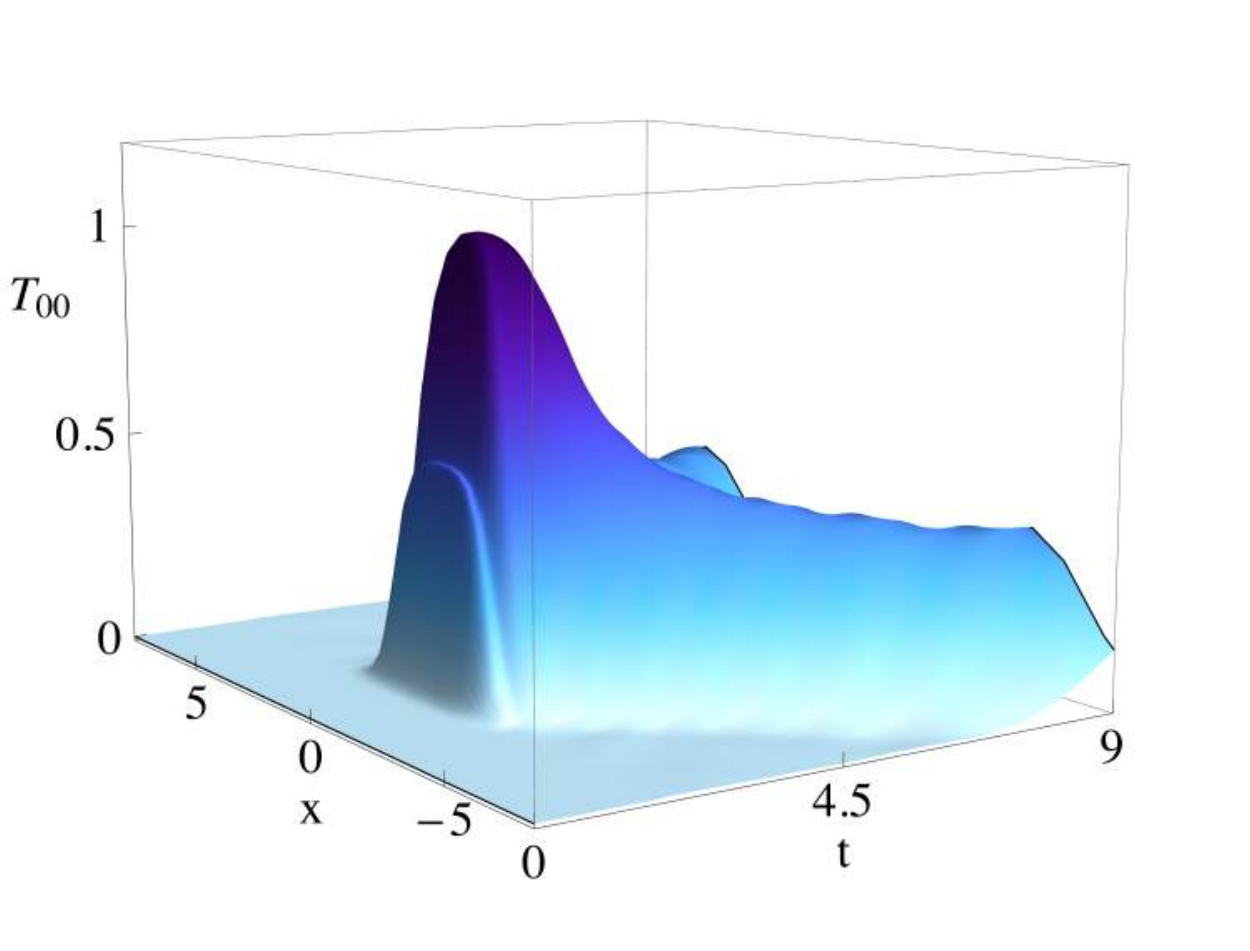}
\subcaption{$T_{00}(t,x)$}
\label{fig_energy}
\end{subfigure}
\caption{Evolution profile of the (a) radial shift $\lambda(x,t)$, and (b)  $T_{00}(x,t)$ component of the stress tensor, for 
$\alpha=0.5$, $M = 0.1$, $\sigma=2$. The field $\lambda$ determines the evolution of the entropy in our solution.}
\label{fig:solution1}
\end{figure*}

The evolution of the spacetime following our quench is fairly simple. The injection of local excitation results in hydrodynamical evolution almost from the very beginning (cf., \cite{Chesler:2008hg,Bhattacharyya:2009uu} for analogous statements with spatial homogeneity). Since our perturbation excites the sound mode of the system, we have the initial energy-momentum perturbation dispersing at the speed of sound. The presence of shear viscosity results in entropy production, manifested in the solution by the local growth of the horizon area element.

\begin{figure}[hb] 
\centering
    \includegraphics[width=8cm]{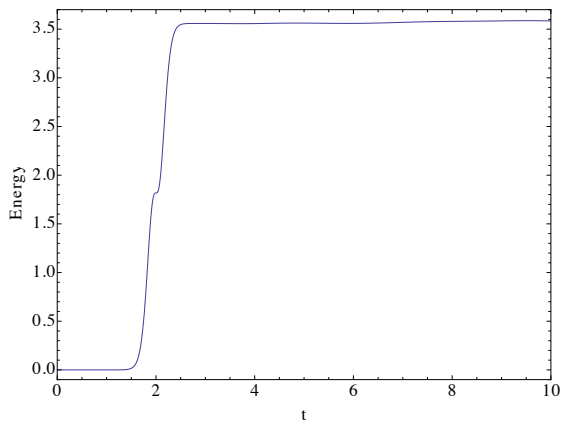}
    \caption[]{Evolution of the total energy on the boundary $E = \int T_{00} \,dx$ after a quench described by parameters $\alpha=0.5$, $M = 0.1$, $\sigma=2$.}
\label{fig_total}
\end{figure}

In  Fig.~\ref{fig_entropy}  we display the spatial and temporal profile of the function $\lambda(x,t)$, related to the area element of the horizon. We see that the initial perturbation indeed results in entropy production, as expected. Curiously, the initial perturbation splits to two localized perturbations after some time; those follow the expected hydrodynamic evolution.  Fig.~\ref{fig_energy} shows the equivalent evolution of the energy density for the same set of parameters. Finally,  Fig.~\ref{fig_total} shows that following the conclusion of the quench the total energy is conserved. These features verify the intuitive picture of hydrodynamical evolution following a local excitation of the system.

To quantify the entropy production, we can monitor the growth of  the area of the apparent horizon as a function of time. In order to express the result in physical units, we need to convert from the natural time scale on the horizon to the time measured in the boundary. Recall that our 
solutions for the metric components are obtained on a slice of constant ingoing time coordinate $t$. We could, following \cite{Bhattacharyya:2008xc}, map the horizon data along ingoing null geodesics to the boundary. We will refrain from doing so explicitly and instead work directly in the chosen coordinates leaving implicit this translation.\footnote{ We also note that $\lambda(t,x)$ is defined on a constant ingoing time slice, and as such the radial shifts affect the horizon ``instantaneously" rather than causally.} 
\begin{figure}[h!] 
\centering
    \includegraphics[width=11cm]{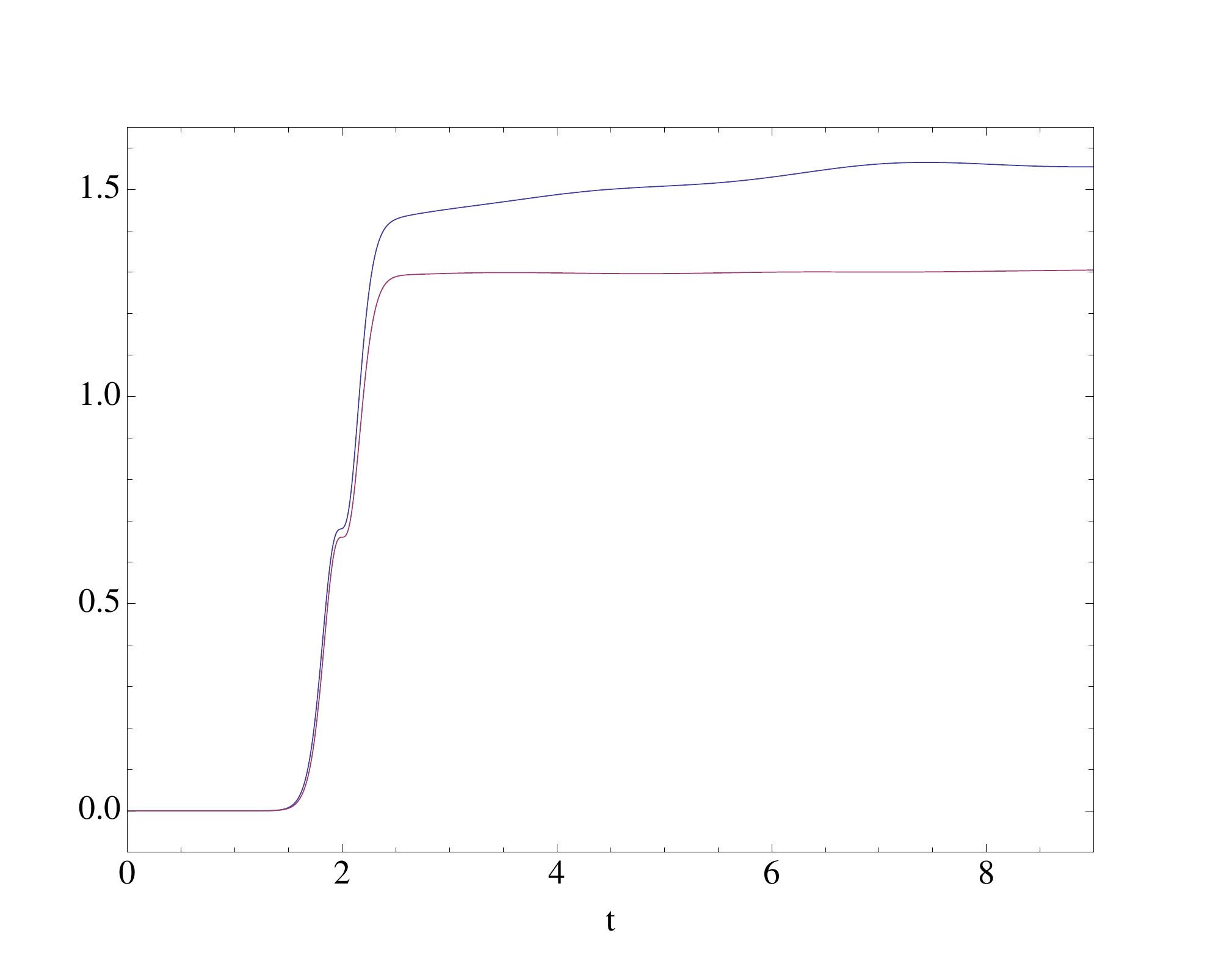}
    \caption[]{I The growth of the apparent horizon (in blue) as a function of boundary time, for $\alpha = 0.3$ and $M = 0.1$. We also overlay the plot for the total energy $\int \,T_{00} dx$ produced  by quenching the system in red for direct comparison.}
\label{fig_horizon}
\end{figure}

Using the induced metric $h_{ab}$ on a constant $t$ slice we obtain the area element on the horizon which can be integrated directly. Since the naive answer is infinite, we regulate it by removing the contribution from the initial equilibrium state (i.e. subtract off the static Schwarzchild-AdS answer) to obtain:
\begin{equation}
 \Delta \text{Area}_h=  L_y \int_{r=r_h}\left(  \Sigma^2 \sqrt{1 + 2 \lambda^\prime e^{-2\chi - B}} - r_h^2 \right) \;dx 
  \end{equation}

The numerical results are expressed in Fig.~\ref{fig_horizon}, where we also show the total energy for comparison. Notice the striking resemblance of the horizon's area evolution with that of the total energy injected into the system by the quenching scalar field. This seems to indicate that the growth of the horizon is dictated by processes governed by the speed of sound, such as energy and momentum transport. This is indeed the intuition we would have from the hydrodynamic regime of slow variations and it is a reassuring check of the set-up that this indeed is upheld.

\subsection{Extremal Surfaces}

\begin{figure}[hb] 
\centering
    \includegraphics[width=9cm]{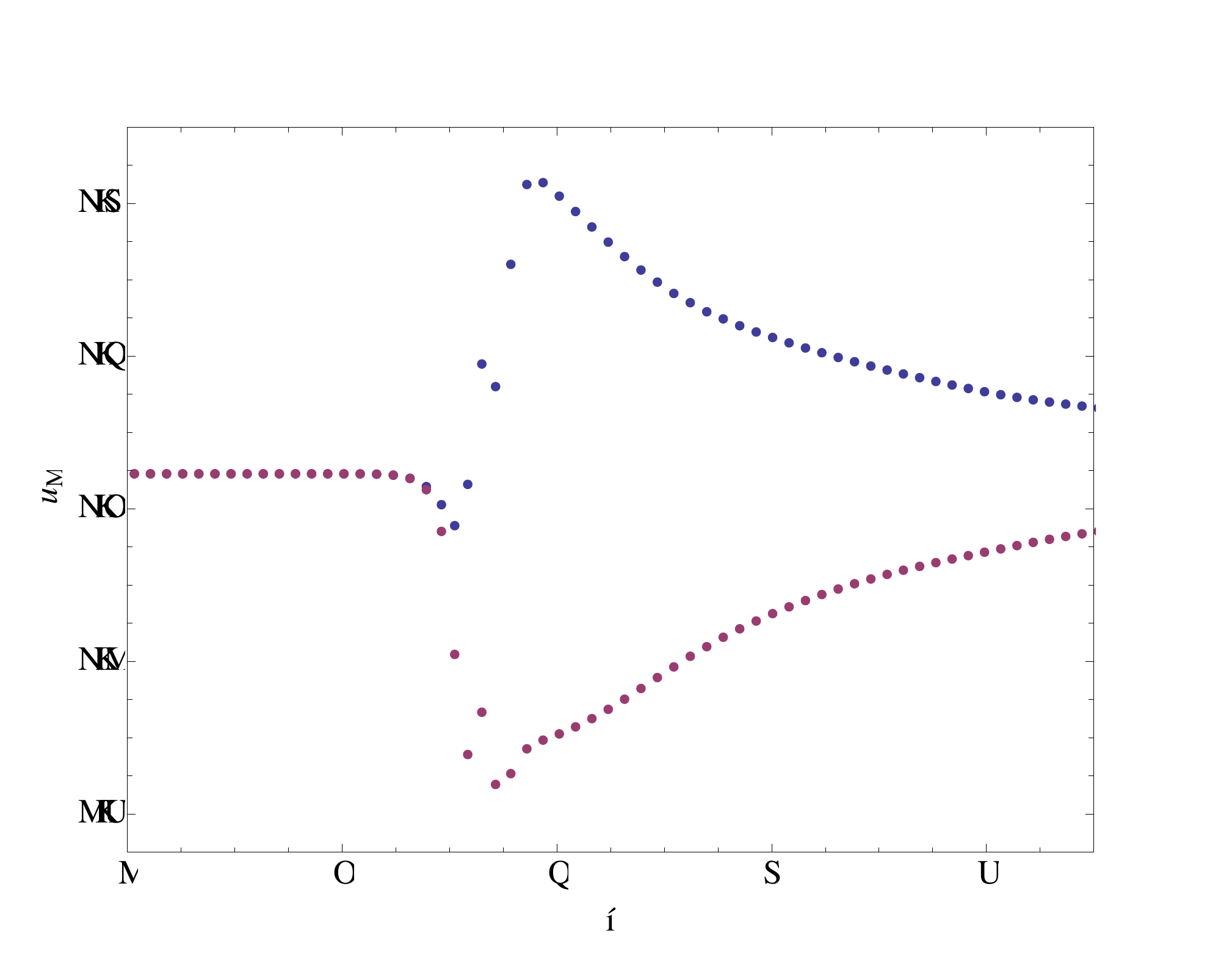}
    \caption[]{Evolution of the geodesics' radial depth for a quench; $\alpha=0.5$, $M=0.1$, $L=0.8$, $\sigma=2$. The blue data points represent the radial depth $u^*$ in the fixed, gauged coordinate system, whereas the red data points represent the ungauged radial depth $U^*$.}
\label{fig_EEradialdepth1}
\end{figure}

Having the solution at hand we can compute the extremal surfaces as described in \S\ref{sec:setupextremal}. In Fig.~\ref{fig_EEradialdepth1} we display the radial depth of the turning point for the extremal surfaces, as function of (boundary) time. Different points correspond to different extremal surfaces, which contribute to entanglement entropy of surfaces of varying lengths. We have plotted the radial depth both in the computational coordinate (in which the horizon is at fixed radial distance) and in coordinates in which the horizon grows. 

Since our calculational domain ends at the apparent horizon, we cannot probe extremal surfaces that extend past into the trapped region. These are known to exist in various explicit simulations (cf., \cite{Hubeny:2013dea} for a comprehensive survey in Vaidya-AdS spacetimes). Pragmatically, this restricts our attention to small regions $\mathcal{A}$.  We will nevertheless see that despite this restriction we can still extract interesting physical features of $S_\mathcal{A}$ using surfaces that lie outside the apparent horizon.

One of the interesting features to notice from Fig.~\ref{fig_EEradialdepth1} is that the geodesics never go beyond their initial depth in the bulk when we consider their position in the \textit{ungauged} radial coordinate, i.e., where the radial depth is 
\begin{align}
 U^* \equiv \frac{u^*}{1 + \lambda(t^*,x=0) u^*} \,, 
\end{align}
with $u^* = 1/r^*$ being the radial position of the tip in the coordinate system where the apparent horizon is at a  fixed coordinate locus.

\section{Propagation of Entanglement Entropy}
\label{sec:results}
%

Armed with the numerical results for the spacetime geometry and the extremal surfaces therein, we are now in a position to extract some physical lessons for the evolution of entanglement entropy following a local quench.  We restrict our attention to regions $\mathcal{A}$ centered around the source of the initial excitation which is taken to be w.l.o.g. at $x=0$. We will examine the behaviour of $\Delta S_\mathcal{A}$ as a function of the width $L$ of the strip and time $t$ after the quench.

We note that the region of parameter space that we can explore numerically is limited. The amplitude $\alpha$ of the scalar field cannot be too large, otherwise the time-evolution of the quench solution does not converge. Similarly, the evolution code becomes unstable if the spatial discretization falls below a critical grid size, which has for consequence that we cannot resolve quenches with width $\sigma$ below a certain threshold. The width $L$ of the entangling surface is in turn constrained by the initial values we can pick for $M$, which determines the position of the event horizon of the initial configuration: if $M$ is taken to be large, then we cannot find extremal surfaces that go deep enough in the bulk to probe larger regions $\mathcal{A}$, whereas if $M$ is taken too small, then it becomes increasingly harder to quench the spacetime with a scalar source. We found that using quenches with width $\sigma = 2$, together with $M$ ranging from 0.005 and 0.2 and $\alpha$ between 0.1 and 0.5, yielded interesting results that remained mostly the same, albeit delayed in time, as those with $\sigma$ chosen larger. 

Before proceeding we remind the reader that for regions $\mathcal{A}$ which are much wider than the width of the quench source profile, there is a time delay before the entanglement entropy starts to change. This is consistent with the causal properties one would required of entanglement. Only when the quench can affect both the region and its complement (by being in the past of the entangling surface) would we expect a change in the entanglement for $\mathcal{A}$. This is clearly borne out in our simulations and is used to benchmark that we are on the right track.

\subsection{An Emergent Light-cone}
%

We first note that the entanglement generated by the local quench is linearly dispersing, i.e., it traces an effective light-cone. This is quite reminiscent of the Lieb-Robinson bound \cite{Lieb:1972wy} in non-relativistic theories, where correlations  follow an effective information light-cone. The speed of entanglement propagation is then denoted by $v_E$ below.

The velocity $v_E$ we find is bounded from below. A-priori one might guess whether the lower bound is given by the speed of sound, which is the speed in which the initial pulse spreads, thereby further exciting the system and generating additional entanglement on larger scales. The true speed is however a bit lower, as we shall see, suggesting that the mechanism of entanglement propagation differs from that which drives physical transport of energy and other conserved charges in the system.\footnote{ A-priori this statement statement appears reasonable, since the propagation of energy in the system is governed by the ability of the system to homogenize, which per se is not the same as becoming quantum entangled. There is thus far no clear mechanism for intuiting entanglement transport in quantum field theories, though the attempts of \cite{Casini:2015zua} suggest potentially interesting mechanisms for the same.}

We therefore interpret the velocity  $v_E$ as the speed in which the initial entanglement, generated locally by the quench, propagates in time. The entanglement velocity can be extracted from the emergent light-cone defined along the curve where $\Delta S_\mathcal{A}(t)$ reaches a maximum for every $L$ in the $L-t$ plane. We remark that unlike the results of  \cite{Nozaki:2013wia}, the height of this peak does not remain constant in our setup. Instead, we find that the maximum value of $S_\mathcal{A}(t)$ increases as we increase $L$. 

\begin{figure}[h!] 
\centering
    \includegraphics[width=10.5cm]{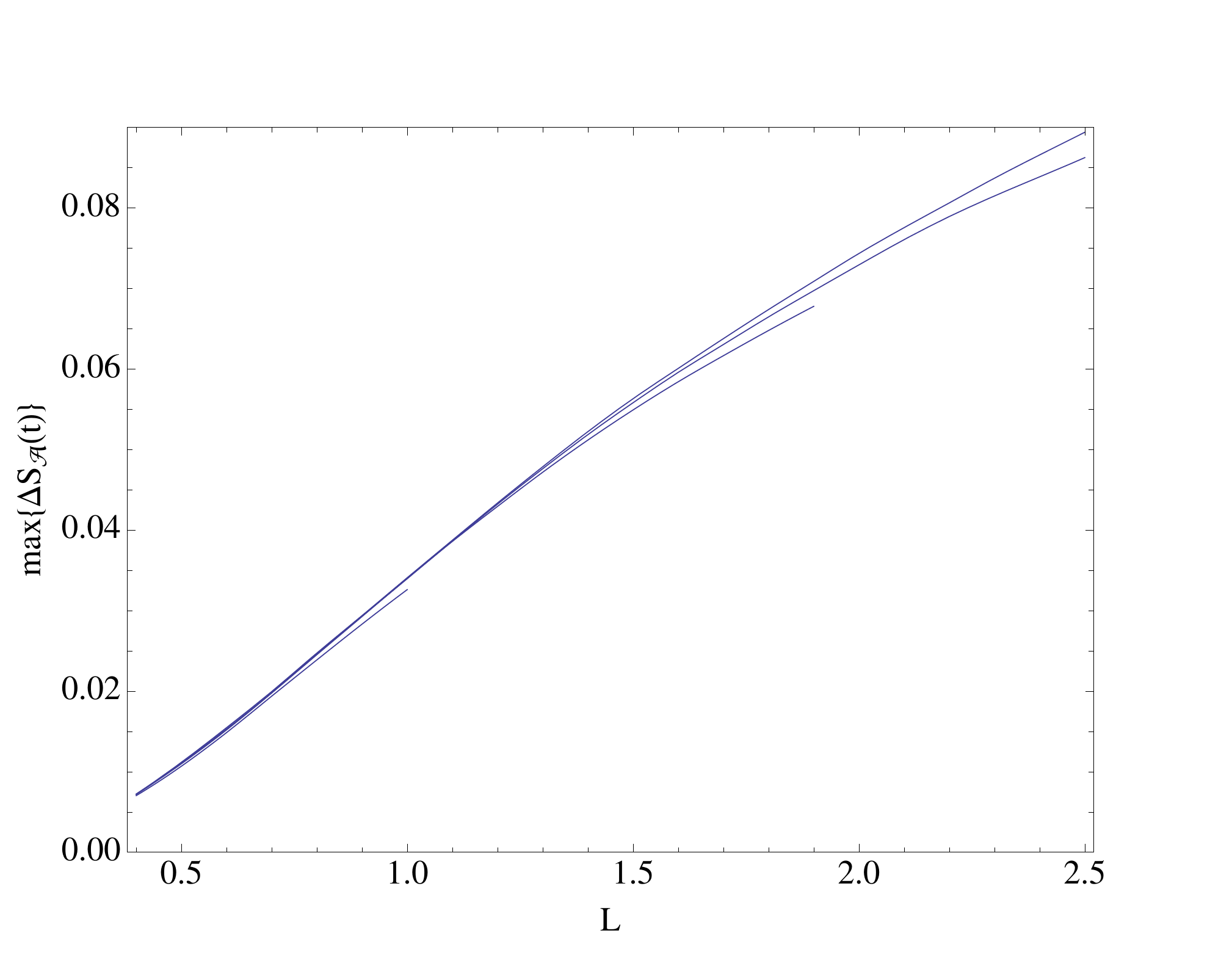}
    \caption[]{Maximum of the entanglement entropy $S_\mathcal{A}(t)$ as a function of $L$, for  a range of masses of the initial black hole $M = \{0.005, 0.01, 0.02, 0.1\}$, starting from the top, and $\alpha = 0.1$. Lowering the temperature (decreasing $M$) slightly increases the maximum of $S_\mathcal{A}(t)$. The same phenomena is observed for $\alpha = 0.2$.}
\label{fig_bckdamplitude}
\end{figure}

This behaviour of the entanglement entropy can be quantified rather explicitly. We find that dependence is strongest when the amplitude of the scalar field is varied. For small sizes $L$, the maximum of $S_\mathcal{A}$ increases linearly with $L$. If we denote the slope of these curves by
 $\mathfrak{s}$, then we find the interesting relation
\begin{equation}
 \frac{\partial}{\partial L} S_\mathcal{A}(L,t_\text{max},\alpha) = \mathfrak{s}(\alpha) \sim \alpha^2 \;\;\; \text{for small/intermediate regions}.
\end{equation}
The actual scaling for the slopes obtained from our numerical data are:
\begin{itemize}
\item $\mathfrak{s}(\alpha) \sim \alpha^{1.92}$ for $\alpha = \{0.1, 0.2, 0.3, 0.4, 0.5\}$ and $M=0.1$ 
\item $\mathfrak{s}(\alpha) \sim \alpha^{2.0043}$ for $\alpha = \{0.05, 0.1, 0.15, 0.2\}$ and $M=0.01$
\end{itemize}
\begin{figure}[h!] 
\centering
    \includegraphics[width=10.6cm]{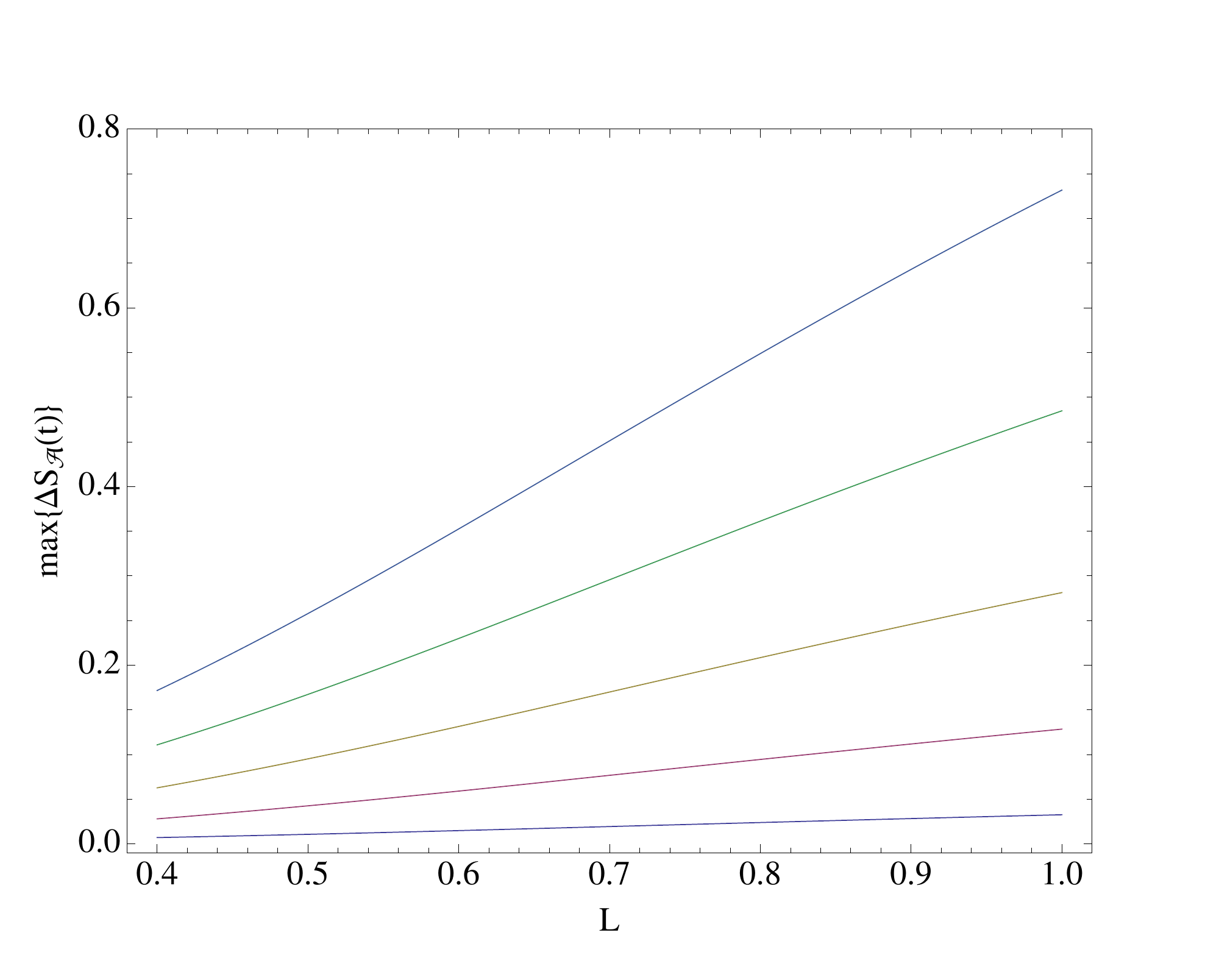}
    \caption[]{Maximum of the entanglement entropy $S_\mathcal{A}(t)$ as a function of $L$, for $\alpha = \{0.1, 0.2, 0.3, 0.4, 0.5\}$ starting from the bottom, and $M=0.1$. The slopes of these curves depend quadratically on the amplitude $\alpha$ of the scalar field.}
\label{fig_scalaramplitude}
\end{figure}
In the first case, the linear behaviour is shown in Fig.~\ref{fig_scalaramplitude}. In the second instance (not pictured), while the linear nature breaks down when $L$ is large, the slopes for small to intermediate regions still depend quadratically on the amplitudes. The dependence on temperature is less interesting. When the temperature $M$ changes, the maximum of the entanglement entropy shifts slightly, as can be seen in Fig.~\ref{fig_bckdamplitude}.

For general values of parameters, the entanglement velocity $v_E$  changes with parameters, always bounded from below by the tsunami velocity \eqref{eq:tsunami}, and above by the speed of light. We do however find two universal results  which we now turn to.

\subsubsection{Universal Behaviour at High Temperature}
\label{sec:uhighT}

\begin{figure}[hb] 
\centering
    \includegraphics[width=9.9cm]{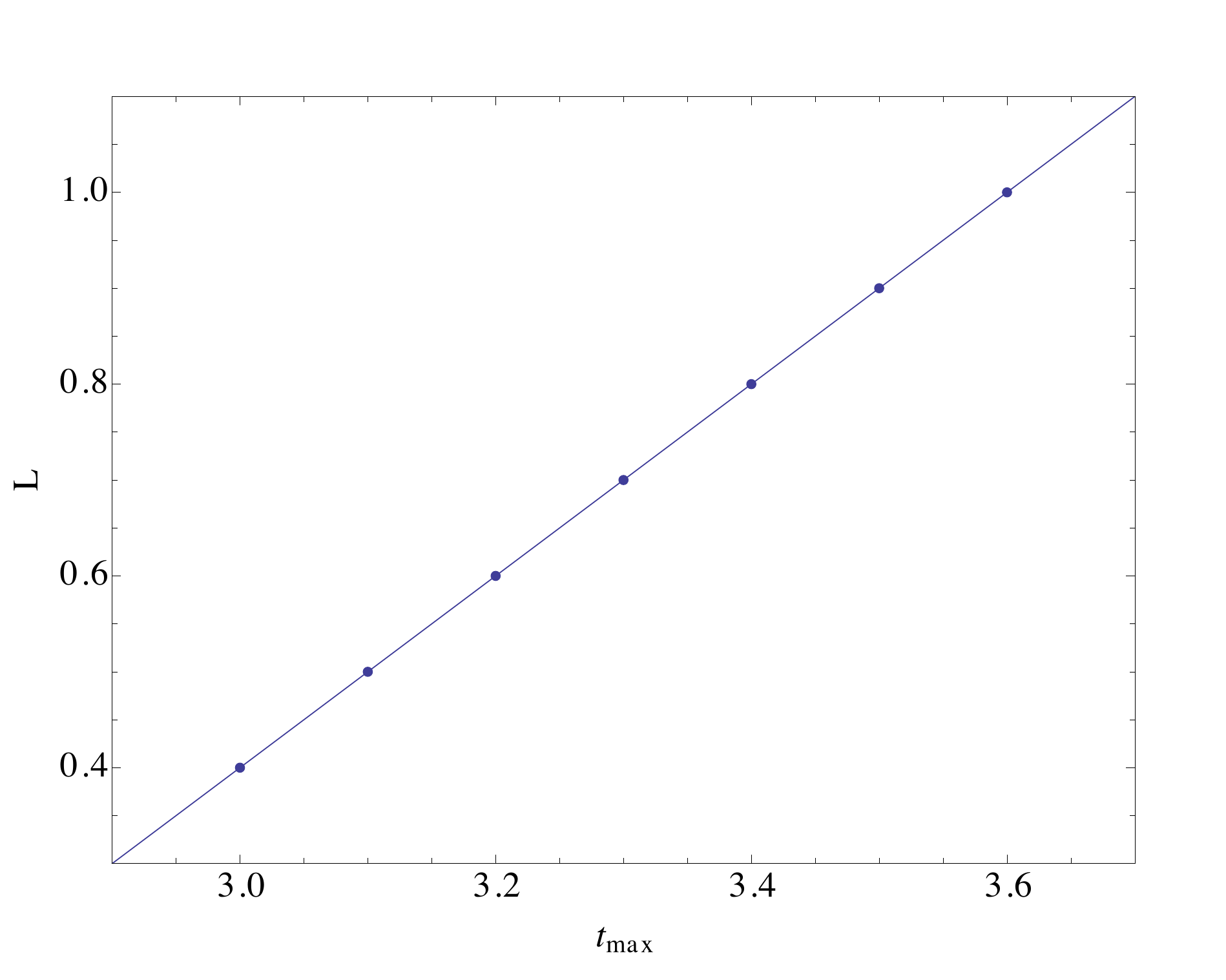}
    \caption[]{Position of the maximum of $S_\mathcal{A}(t)$ in the $L-t$ plane for a quench described by $\alpha=0.5$, $M=0.1$. The light-cone velocity extracted from the slope of this line is $v_\text{LC} = 1$, and is independent of the value of $\alpha$.      }
\label{fig_universal1}
\end{figure}

In the limit of an approximate global quench where the region $\mathcal{A}$ is contained within the local quench, i.e.,
 $L \lesssim \sigma$, and at high temperatures, we find a universal light-cone velocity $v_E = 1$ (to very high accuracy), regardless of the amplitude of driving scalar field (including values well within the non-linear regime)\footnote{ It is worth noting that previous results for global quenches could not have seen this feature since the entanglement entropy saturates for strip geometries.}. This is depicted in Fig.~\ref{fig_universal1}. We note that for some values of parameters, this universal behaviour can be affected by edge effects of the local quench, and is seen for small enough surfaces only.

As we decrease the black hole temperature, the velocity at the small surfaces becomes lower than 1. This confirms  that $v_E=1$ is a high temperature effect only.

\subsubsection{Wide Quench Profiles}

\begin{figure*}[h]
\begin{subfigure}{0.27\textwidth}
   \includegraphics[width=6cm]{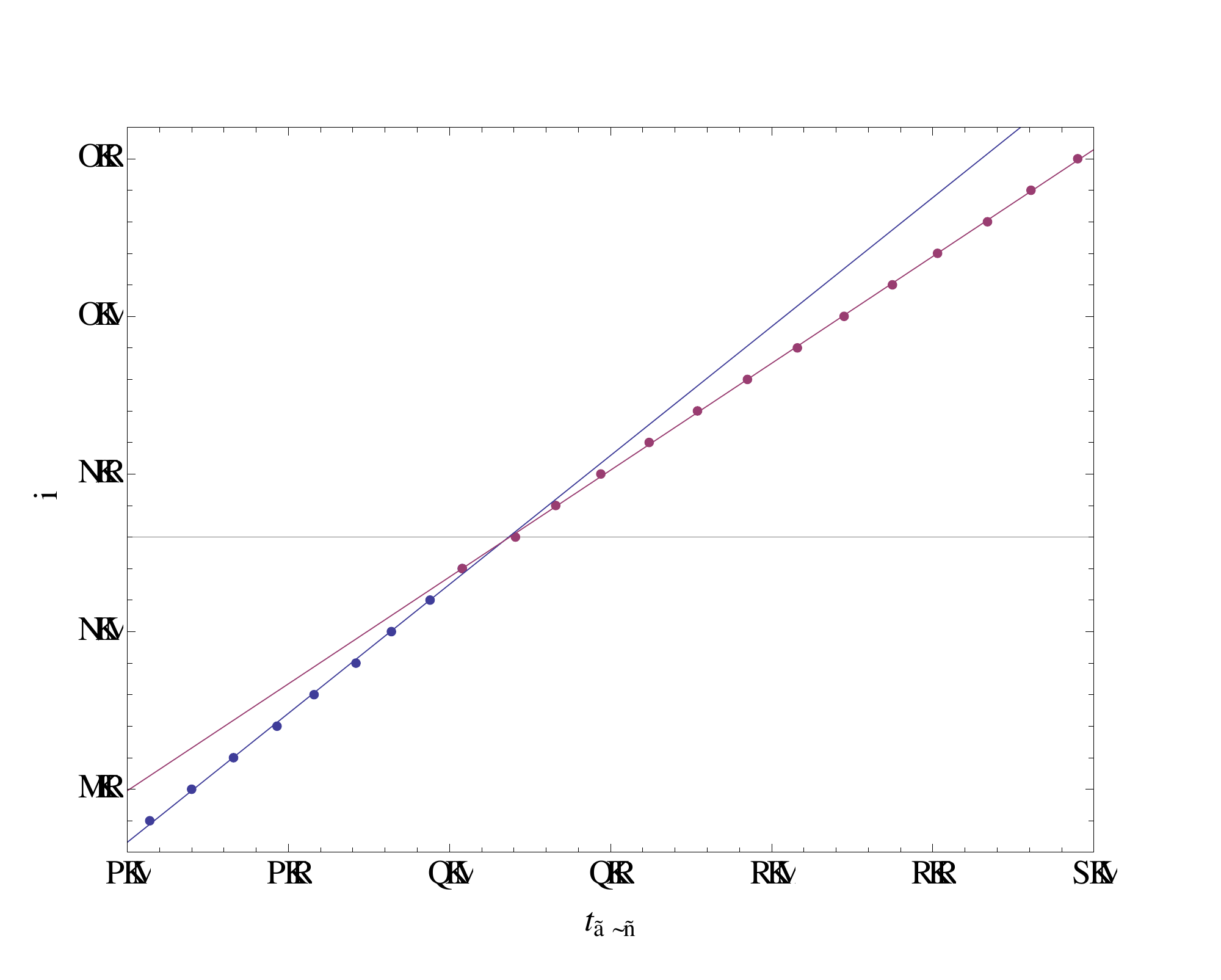}
    \subcaption{$M =0.005,\;  v_E = 0.678 (0.818)$}
\label{fig_change1}
\end{subfigure}
\hfill
\begin{subfigure}{0.28\textwidth}
   \includegraphics[width=6cm]{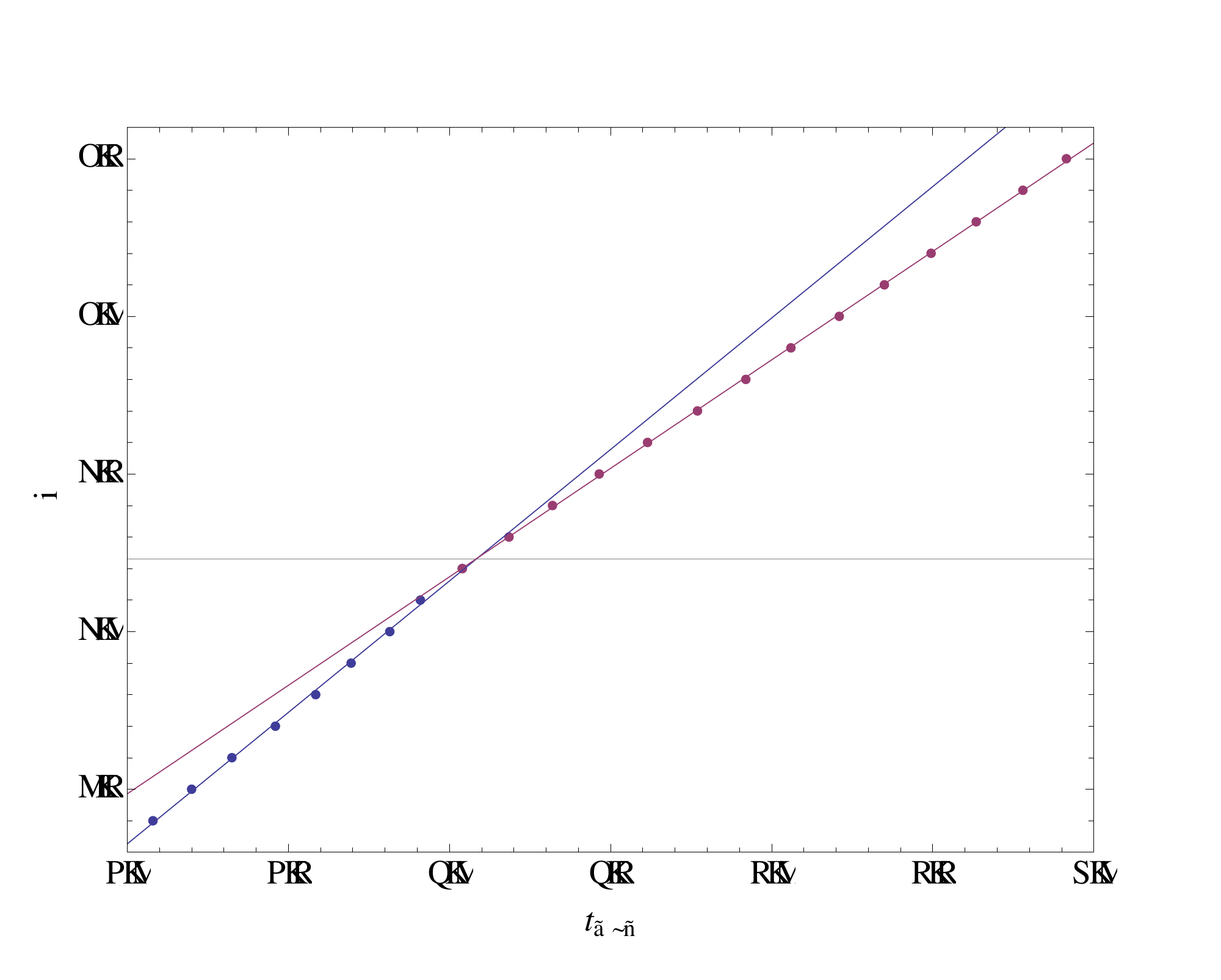}
   \subcaption{$M =0.01,\;  v_E = 0.688 (0.834)$}
\label{fig_change2}
\end{subfigure}
\hfill
\begin{subfigure}{0.27\textwidth}
   \includegraphics[width=6cm]{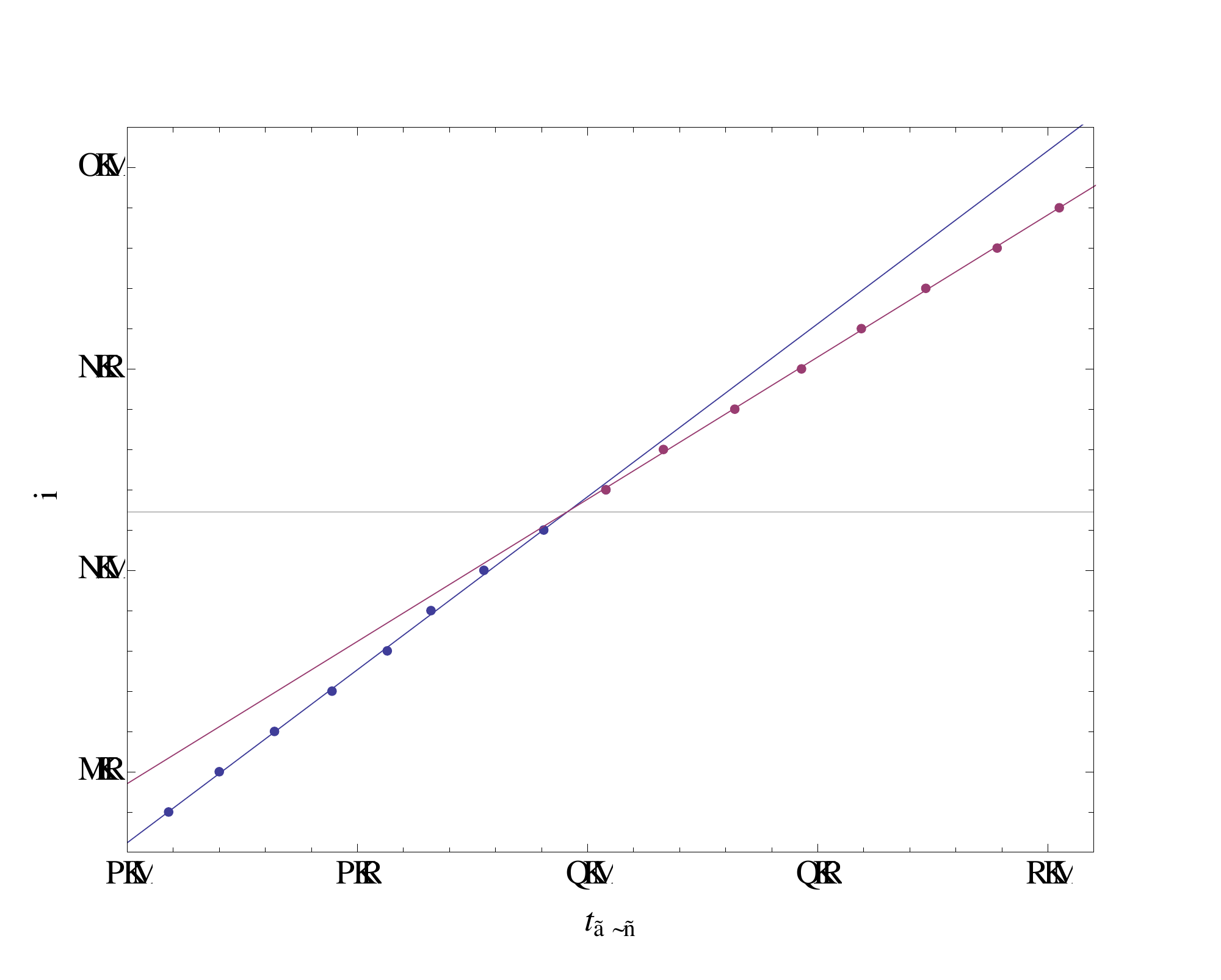}
     \subcaption{$M =0.02, \; v_E = 0.706 (0.859)$}
\label{fig_change3}
\end{subfigure}
\caption{Position of the maximum of $S_\mathcal{A}(t)$ in the $L-t$ plane for a quenches described by $\alpha=0.1$, starting from different initial states parameterized by $M$ shown above. The light-cone velocities for large $L$ for the three scenarios are also indicated, as are the corresponding values for small region sizes (in parenthesis). While we give the values of  the velocity $v_E$ for small regions, this data should be interpreted with care, for we typically find that edge effects contaminate the data, and these slopes should not be taken at face value in the small $L$ regime.}
\label{fig:changes123}
\end{figure*}

An interesting feature of the emergent light-cone is the abrupt change of velocity as the width of the region $\mathcal{A}$, $L$, is increased. When the size of the region $\mathcal{A}$ becomes of the same order as the width of the local quench, the curve traced by the peak of the entanglement entropy goes from one linear regime to another, as shown in Figs.~\ref{fig_change1}, \ref{fig_change2}, and \ref{fig_change3}.

\begin{figure*}[h]
\begin{subfigure}{0.49\textwidth}
  \includegraphics[width=8cm]{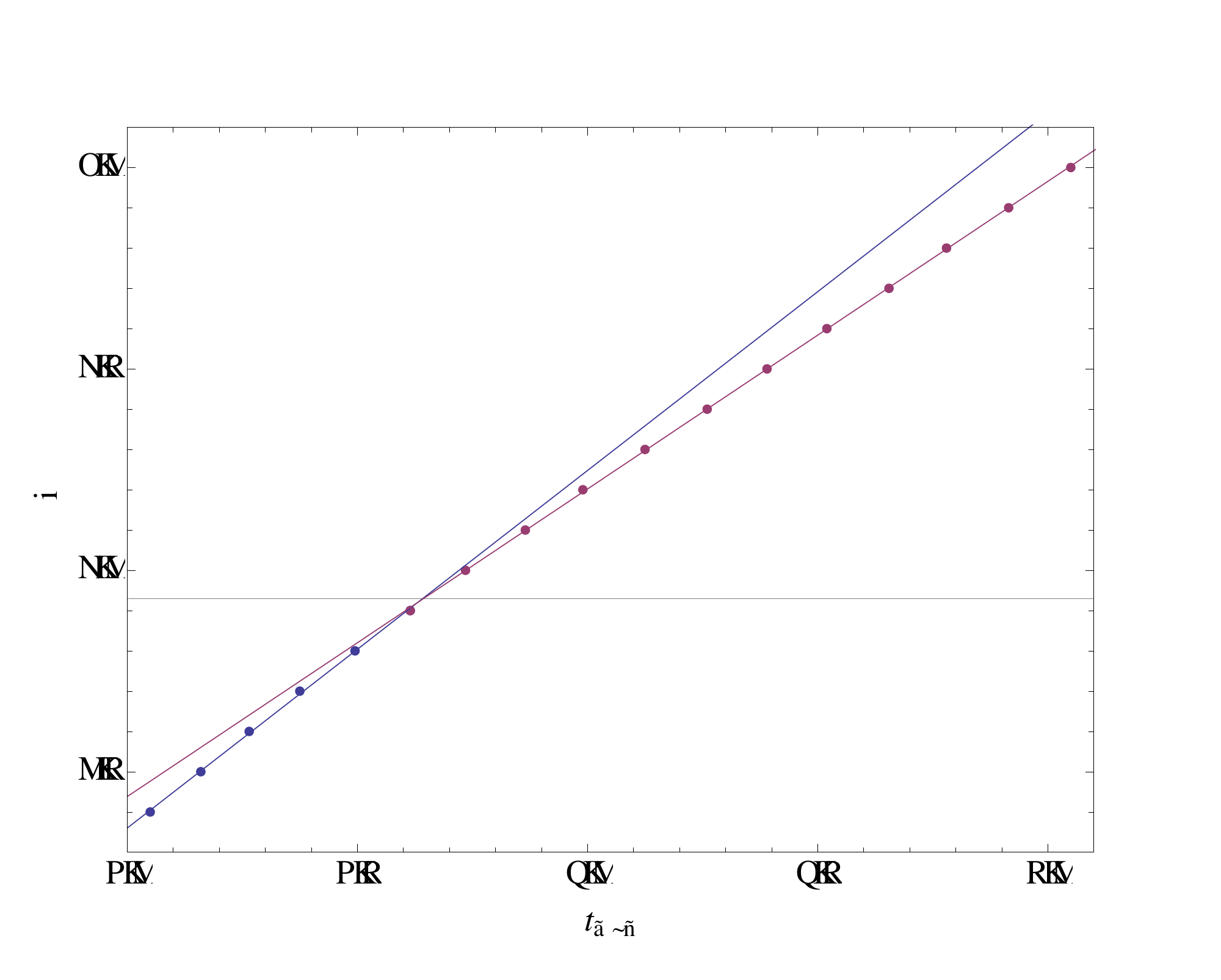}
    \subcaption{$M =0.01,\; v_E = 0.764 (0.887)$}
\label{fig_change4}
\end{subfigure}
\hfill
\begin{subfigure}{0.49\textwidth}
      \includegraphics[width=8cm]{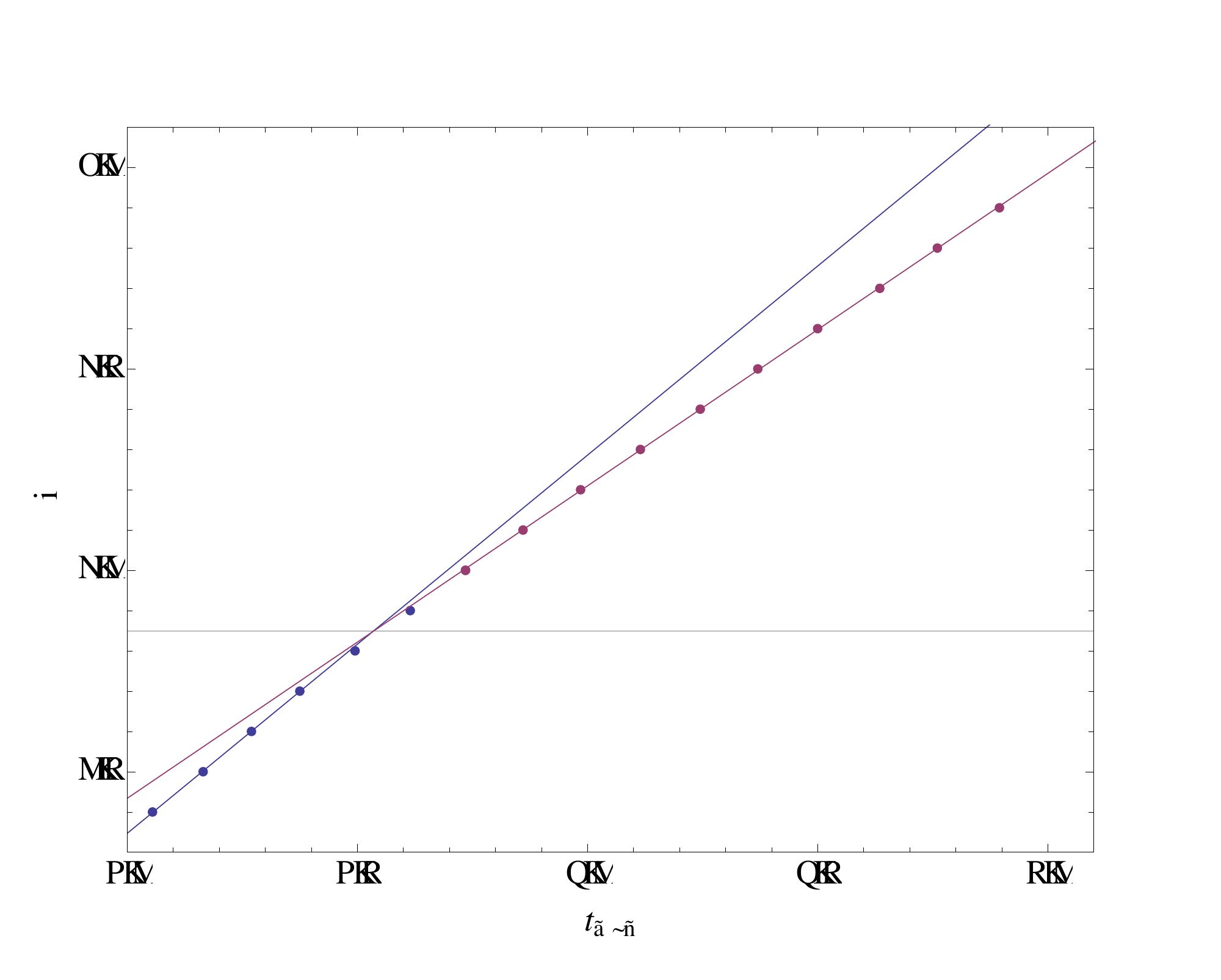}
   \subcaption{$M =0.02, \;  v_E = 0.776 (0.934)$}
\label{fig_change5}
\end{subfigure}
\caption{Position of the maximum of $S_\mathcal{A}(t)$ in the $L-t$ plane for a quenches described by $\alpha=0.2$, starting from different initial states parameterized by $M$ shown above. The light-cone velocities for large $L$ for the three scenarios are also indicated.  Conventions are the same as in Fig.~\ref{fig:changes123}.
}
\label{fig:changes45}
\end{figure*}

Interestingly, for the first two data sets (for which $\alpha=0.1$, $M=\{0.005, 0.01, 0.02\}$, $\sigma=2$), the light-cone velocities of $v_E= \{ 0.678, 0.688, 0.706\}$ are very close to the tsunami velocity of a Schwarzschild-\AdS{d+1} black hole found in \cite{Liu:2013iza}, given by
\begin{equation}
v_E^{*}(3) = \frac{(\eta - 1)^{\frac{1}{2} (\eta - 1)}}{\eta^{\frac{1}{2} \eta}} \Bigg \vert_{d=3} = \frac{\sqrt{3}}{2^\frac{4}{3}} = 0.687, \;\;\; \text{with} \;\; \eta = \frac{2(d-1)}{d}.
\label{eq:tsunami}
\end{equation}	

We note that temperature does not seem to have an effect on $v_E$, which is consistent with the above formula. For these parameters, the evolution is described by linear response to good approximation, and in that regime the tsunami velocity seems to capture the spatial propagation of entanglement to very good accuracy.

This behaviour should be anticipated on physical grounds. When the region $\mathcal{A}$ is completely immersed in the quench source, we are back to the case where we may approximately think of the situation as a global quench problem. The fact that the source is not homogeneous in $\mathcal{A}^c$ is irrelevant because all that matters is that the excitations produced by the quench are in the causal past of  the entangling surface $\partial \mathcal{A}$. With this in mind we immediately anticipate that the results for the Vaidya quench explore in \cite{Liu:2013iza,Liu:2013qca} should apply and one see a linear growth with the tsunami velocity.

The story of the local quench however should be a lot richer than the homogeneous global quench. For one, we can encounter an interplay between the size of $\mathcal{A} $ and the width of the pulse. We also expect that the non-linearities of gravity will play a role as we try to increase the amplitude. Indeed we see that  velocity $v_E$ increases  as we increase the strength of the non-linearities in the bulk evolution -- this is illustrated in Figs.~\ref{fig_change4} and \ref{fig_change5} (where the scalar field amplitude was doubled from 0.1 to 0.2). This goes against the idea of the tsunami velocity as an upper bound on the speed propagation of the entanglement propagation, at least when that evolution is spatially resolved.
Coupled with the earlier observation regarding the upper bound on $v_E \leq 1$, we find it natural to conjecture that 
\begin{equation}
v_E^*(3) = 0.687 \leq v_E \leq 1
\label{eq:vebdd}
\end{equation}	
The details of deviation from the two extreme limits appear to depend on various effects which we have not yet disentangled. While the upper bound follows form causality, it is unclear at present whether the tsunami velocity encountered (herein and before) is a fundamental bound on information processing in strongly coupled systems. It would be interesting to come up with a model which allows us to explore the different propagation velocities perhaps along the lines of \cite{Casini:2015zua}.

\subsection{Entanglement Decay}
%
%
The process of return to equilibrium is characterized by universal behaviour and critical exponents. Therefore, an interesting quantity in our model is the decay of the entanglement entropy after it has reached a local maximum. To our knowledge this is the first time this decay has been calculated in either  holographic theories or in higher dimensional conformal field theories.

\begin{figure}[hb] 
\centering
    \includegraphics[width=10cm]{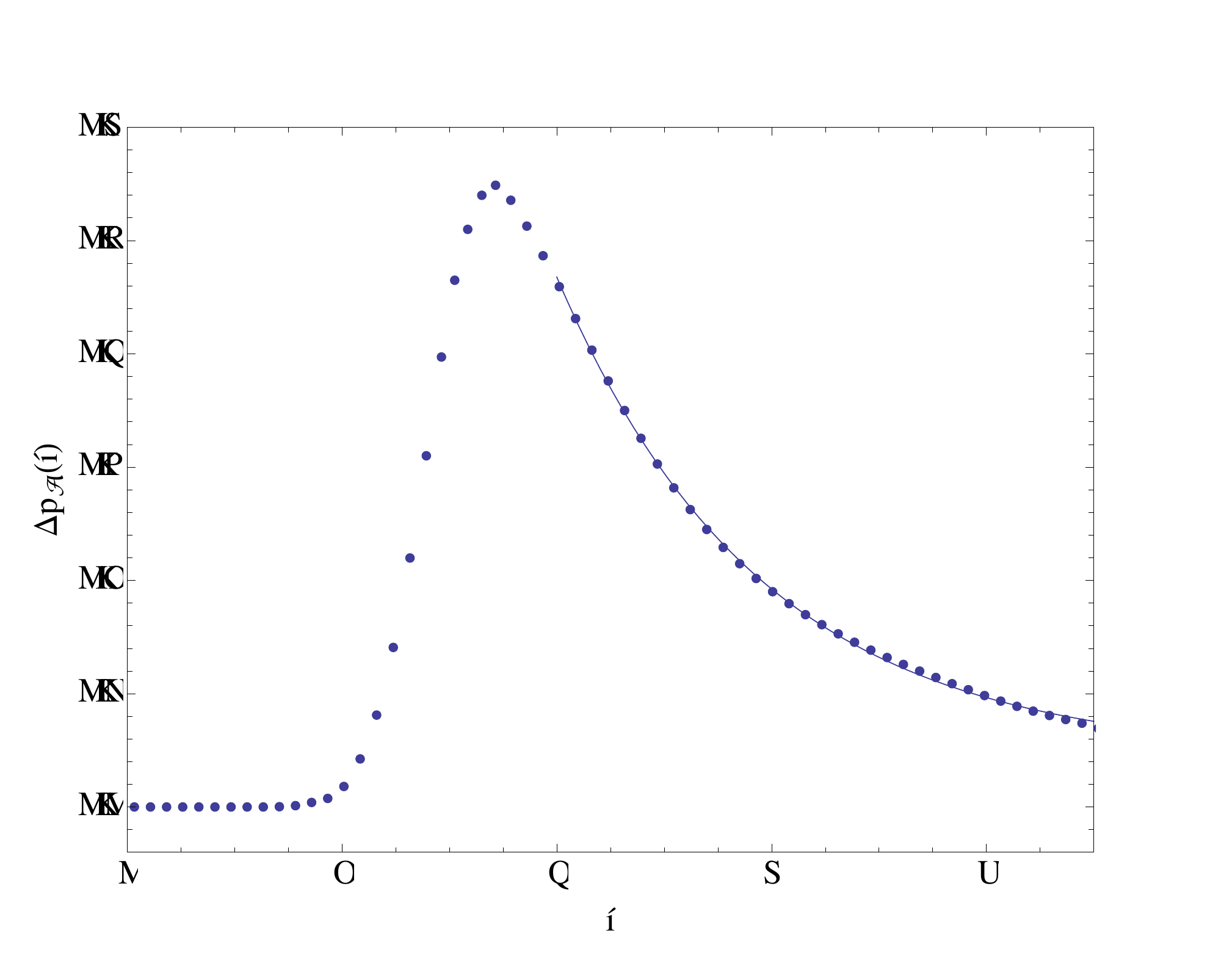}
    \caption[]{Exponential decay of the entanglement entropy evolution at late times; $\alpha=0.5$, $M=0.1$, $L=0.8$.  The fit parameters for the particular choice of quench parameters turns out to be  $a_1 = 2.5335$, $a_2 = 0.5277$, $a_3 = 0.6049$, and $a_4 = 0.0454$. Note that we evolve the solution for late but finite time, which explains why $a_4 \neq 0$. In the infinite time limit we expect $a_4=0$.}
\label{fig_EEdecay1}
\end{figure}

From our numerical data we find that the profile for the decay is best fitted by an exponential damping
\begin{equation}
\Delta S_\mathcal{A} (t) \sim a_1 e^{-a_2 (t - a_3)} + a_4 \,, 
\end{equation}	
where the parameters $a_i$ depend on the specifics of the sources chosen to implement the quench protocol. In  Fig.~\ref{fig_EEdecay1} we depict the behaviour for  a particular simulation (parameters in the caption). Note also the time delay in the initial growth, which illustrates the causality feature discussed earlier.

It is interesting to contrast our result for the exponential return to equilibrium against a more slow return seen in some spin chain models. For instance, in \cite{Eisler:2007aa} the authors study free electrons in a half-filled chain and determined the growth and decay of the entanglement entropy after a local quench. In that set-up they find a very slow return to the unperturbed value. In two dimensions the decay is characterized by $ S_\mathcal{A} (t) \sim \frac{ a_1 \log(t) + a_2}{t}$ as  $t \rightarrow \infty$. The parameters $a_1,a_2$.are again obtained by fitting and depend on the specific details of the quench.

It is somewhat intriguing that the holographic computations relax much faster. This is reminiscent of features of scrambling in black hole physics, which we comment on in our discussion \S\ref{sec:discuss}.

\section{Conclusions and Future Directions}
\label{sec:discuss}

The main focus of the present paper was to describe the dynamics of the holographic entanglement entropy following a local quench. While this problem has been studied in the past using various known exact solutions to model the quench, we have carried out a full numerical simulation of Einstein's equations in the presence of a perturbing external source on the boundary of AdS. Given the explicit numerical solution to the quench geometry, we can study the dynamics of entanglement entropy by exploring the behaviour of extremal surfaces that are anchored on the boundary.

The upshot of our analysis was a clear signal that entanglement entropy disperses linearly, in a manner reminiscent of the Lieb-Robinson light-cone. The dispersion velocity appears to depend on the details of the quench, though we were able to bound the result between two interesting bounds that have been discussed in the literature earlier. On the one hand we found that for wide quench profile, the propagation speed saturated a putative lower bound, given by the entanglement tsunami velocity obtained by \cite{Liu:2013iza} in the context of global quenches (modeled using the Vaidya-AdS spacetime). On the other hand well localized quenches appear to propagate entanglement at the speed of light. It is rather curious that we have results very similar to the Vaidya-AdS quench, for the geometry we construct is not the same. This lends support to thesis of \cite{Liu:2013iza,Liu:2013qca} that the holographic tsunami velocity ought to be a generic phenomenon. 

The second aspect of holographic entanglement entropy which is interesting in our study is the rather rapid reversion of result to the equilibrium value. In various simulations we have tested, the reversion is exponentially fast, in contrast to the much slower logarithmic decay seen in spin models. This suggests again, as has been suspected in the past, that black holes are very efficient at information processing, cf., \cite{Hayden:2007cs,Sekino:2008he}.

There are many other interesting areas for further investigation. It would be interesting to study other quench protocols and other theories, including massive models, primarily to extract a more detailed dependence of the entanglement velocity and the rate of equilibration. A particularly interesting direction is the study of (global and local) quenches past critical points, generalizing the results of  \cite{Cardy:2015xaa} to higher dimensions.  It would also be interesting to study other non-local measures besides the entanglement entropy, which are more sensitive to the spatial structure of entanglement in quantum field theory, and to the differences between strongly coupled holographic CFTs and CFTs of small central charge. In particular, the mutual information of disjoint intervals would be interesting to calculate in our setup for local quenches.   Finally, one can make a direct connection to the study of entanglement entropy following a local quench in two-dimensional CFTs, for which we have analytic results to explain the behaviour at large central charge \cite{Asplund:2014coa}.  We hope to report on these results in the near future  \cite{Rangamani:2016aa}.

\section*{Acknowledgements}

We have benefited from conversations with  Tom Faulkner, Tom Hartman, Veronika Hubeny,  Hong Liu, Andy Lucas, Josephine Suh, and Mark Van Raamsdonk.  M.~Rangamani would like to thank Galileo Galilei Institute for Theoretical Physics for hospitality during the concluding stages of this work.

M.~Rangamani was supported in part by the ERC Consolidator Grant Agreement ERC-2013-CoG-615443: SPiN and by the FQXi  grant ``Measures of Holographic Information"  (FQXi-RFP3-1334).  AVE and M.~Rozali are supported by a discovery grant from NSERC of Canada. 

\pagebreak
\appendix
\section{Apparent Horizons}
\label{sec:apparenthorizon}

In  \S\ref{sec:setupquench}, we used the residual radial reparametrization freedom of the metric \eqref{eq:rrepar} 
to fix the coordinate location of the black hole's apparent horizon.  Here we provide some details on the process we used.

The notion of apparent horizon depends on the existence of trapped surfaces, which in turn depend on a chosen foliation. Given a spacelike surface $\mathcal{S}$, a trapped surface on $\mathcal{S}$ corresponds to the region where both ingoing and outgoing future-directed null geodesic congruences orthogonal to $\mathcal{S}$ have non-positive expansions. The apparent horizon is then defined as the boundary of this trapped region, on which the geodesic congruences have vanishing expansions. 

In our case, a spacelike surface can be parametrized by the two orthogonal vector fields spanning the $x$ and $y$ direction:
\begin{equation}
e_x^M = ( 0,0,1,0), \;\;\;\;\; e_y^M = (0,0,0,1). 
\end{equation}	

We now construct future-directed null geodesic congruences orthogonal to both $e_x$ and $e_y$. Ingoing geodesic congruences can be parametrized by the tangent null vector field $k^M = (0,-1,0,0)$, whereas outgoing geodesic congruences have 
\begin{equation}
 N^M = e^{-2 \chi} \left(1,\; A+e^{-2 \chi}\, \frac{e^{-B}\,F_x^2 }{2 \,\Sigma^2}, \; \frac{e^{-B}\,F_x}{\Sigma^2},\; 0
  \right). 
 \label{aux_null} 
\end{equation}	
The normalization is chosen such that $g_{MN}\, k^M\, N^N = -1$. Since we are interested in the rate of change of the cross-sectional area of null geodesic congruences along their transverse directions, we need to define the transverse metric
\begin{equation}
h_{MN} = g_{MN} + k_M N_N + N_M k_N . 
\label{}
\end{equation}	

With this in hand, we can calculate the expansion $\theta \equiv h^{MN} \nabla_M k_N$.\footnote{ The expansion for the ingoing geodesic congruences along $n$ is always negative, so we need only worry about the congruences along $k$.} This yields a condition on the dynamics of the field $\Sigma$:
\begin{align}
 \left[ d_+\Sigma - \frac{e^{-B}}{2\,\Sigma} \left( F_x \,\partial_x B - \partial_x F_x - e^{-2 \chi} \,F_x^2 \,
 \frac{\partial_r \Sigma}{\Sigma} \right) \right]_{r=r_h} = 0 \,, 
\label{appcond} 
\end{align}
where $d_+ \equiv \partial_t + A \; \partial_r$.  In addition, taking a time derivative of this relation yields a stationarity condition that ensures that the horizon condition holds for all times. One can show that the resulting constraint can be expressed as a second order spatial ODE that determines the value of $A(r,t,x)$ at the apparent horizon.

\section{Numerical Details: Integration Strategy and Boundary Conditions}
\label{sec:boundaryconditions}

We use the characteristic formulation of Einstein's equations resulting from the null slicing of spacetime outlined in \cite{Chesler:2013lia} (see also \cite{Balasubramanian:2013yqa}) to numerically integrate our solution. The clever idea behind this scheme is that both $A(r,t,x)$ and time derivatives disappear completely from the equations of motion if we replace the latter with $d_+  = \partial_t + A \; \partial_r$, the directional derivative along outgoing null geodesics. The equations thereby obtained reduce to a set of nested radial ODEs that is much easier to tame than Einstein's equations in all their glory. 

For numerical purposes, we need to change variables to $u = 1/r$ to make the domain compact, and redefine the fields appearing in the metric by subtracting the known divergent pieces as $u \rightarrow 0$. We do so as follows: 
\begin{equation}  
\begin{split}
\Phi(u,t,x) \equiv& \;\; \phi(u,t,x) u, \\
\Sigma(u,t,x) \equiv& \;\; \frac{1 + \lambda(t,x) u}{u}  - \frac{1}{4} \phi(u,t,x)^2  u,  \\
B(u,t,x) \equiv& \;\; b(u,t,x) u^3, \\
\chi(u,t,x) \equiv& \;\; c(u,t,x) u^3, \\
 F_x(u,t,x) \equiv& \;\;- \partial_x \lambda(t,x) + f_x(u,t,x), \\
 d_+ \Sigma(u,t,x) \equiv& \;\; \frac{(1 + \lambda(t,x)u )^2}{2 u^2} + \tilde{\Sigma}(u,t,x), \\
 d_+ \Phi(u,t,x) \equiv& \;\; \tilde{\Phi}(u,t,x), \\
 d_+ B(u,t,x) \equiv& \;\;  -\frac{3}{2}b(u,t,x) u^2 + \left(  \tilde{B}(u,t,x) 
 	+ \frac{1}{2} \partial_u b(u,t,x) \right) u^3, \\
 A(u,t,x) \equiv& \;\; \frac{(1+\lambda(t,x) u)^2}{2 u^2} + a(u,t,x) 
\end{split}
\end{equation}
We remark that the presence of a scalar source forces us to shift the field $\Sigma$ by an appropriate function of $\phi$ so that it satisfies our asymptotic analysis. 

Given $\phi$, $\lambda$, and $b$, we can proceed to solve for $c$, $f_x$, $\tilde{\Sigma}$, $\tilde{\Phi}$, and $\tilde{B}$ -- in that order. The prescription to find appropriate boundary conditions for these new fields is to expand the equations of motion near $u=0$ and verify their agreement with the asymptotic analysis conducted in  
\S\ref{sec:asympt}. We have, after an appropriate normalization of the ODEs:
\begin{equation}   
\begin{split}
\frac{u}{3} \; \partial_u c(u,t,x) + c(u,t,x) =& \; g_c(t,x) \;\;\;  \Rightarrow \;\;\; c(u,t,x) 
	= \frac{C_c(t,x)}{u^3} + g_c(t,x) , \\ 
-\frac{u^2}{2} \; \partial_u^2 f_x(u,t,x) - u\;  \partial_u f_x(u,t,x) \; +& \; f_x(u,t,x) = \; g_{f_x}(t,x)  \\
 \Rightarrow \;\;\; f_x(u,t,x) =& \; \frac{C_{f_x}^{(1)}(t,x)}{u^2}  + C_{f_x}^{(2)}(t,x) \; u + g_{f_x}(t,x), \\
-u \; \partial_u \tilde{\Sigma}(u,t,x) + \tilde{\Sigma}(u,t,x) =& \; g_{\tilde{\Sigma}}(t,x) \;\;\;  \Rightarrow \;\;\;  \tilde{\Sigma}(u,t,x) = C_{\tilde{\Sigma}} (t,x) \; u+ g_{\tilde{\Sigma}}(t,x),  \\
-u \; \partial_u \tilde{\Phi}(u,t,x) + \tilde{\Phi}(u,t,x) =& \; g_{\tilde{\Phi}}(t,x) \;\;\;  \Rightarrow \;\;\;  \tilde{\Phi}(u,t,x) = C_{\tilde{\Phi}} (t,x) \; u+ g_{\tilde{\Phi}}(t,x), \\
\frac{u}{2} \; \partial_u \tilde{B}(u,t,x) + \tilde{B}(u,t,x) =& \; g_{\tilde{B}}(t,x) \;\;\;  \Rightarrow \;\;\;  \tilde{B}(u,t,x) = \frac{C_{\tilde{B}} (t,x)}{u^2} + g_{\tilde{B}}(t,x) \,. \nonumber
\end{split}
\end{equation}         
\begin{itemize}
\item  For $c$, we find that
\begin{equation}
 g_c(t,x) = \frac{1}{6} \,\phi_0(t,x) \,\phi_1(t,x) - \frac{1}{12} \, \lambda(t,x) \, \phi_0(t,x)^2
\end{equation}	
agrees with the asymptotic expansion for $\Sigma$; spectral methods take care of making the non-analytic part vanish: $C_c(t,x) = 0$.

\item For $f_x$, we find that $g_{f_x}(t,x) = \mathcal{O}(u^2)$. As a result, we need to specify $\partial_u f_x(u=0,t,x) = f^{(3)}(t,x)$ as a boundary condition, which in turn determines $C_{f_x}^{(2)}(t,x)$. Again, spectral methods ensure $C_{f_x}^{(1)}(t,x) =0$. 

\item For $\tilde{\Sigma}$, we find
\begin{equation}
 g_{\tilde{\Sigma}}(t,x) = -\frac{1}{8} \phi_0(t,x) ^2 + \mathcal{O}(u)\,,
\end{equation}	
which agrees with the asymptotic expansion for $d_+\Sigma$. However, we do things a bit differently to determine $\tilde{\Sigma}$ on the entire domain \cite{Balasubramanian:2013yqa}. Indeed, in order to ensure that the computational domain remains on a fixed rectangular grid, a condition (\ref{appcond}) on $d_+ \Sigma$ was derived in Appendix \ref{sec:apparenthorizon}. We use this horizon condition as a boundary condition for $\tilde{\Sigma}(u=u_h, t,x)$. As a consistency check, one can verify that
\begin{equation}
\partial_u \tilde{\Sigma}(u=0,t,x) = \frac{1}{2} T_{00}(t,x) - \frac{1}{3} \phi_0(t,x) \phi_1(t,x) - \frac{1}{12} \lambda(t,x) \phi_0(t,x)^2 
\label{}
\end{equation}	
is indeed satisfied at every time step. 

\item For $\tilde{\Phi}$, we find
\begin{equation}
g_{\tilde{\Phi}}(t,x) = -\frac{1}{2} \phi_0(t,x) + \mathcal{O}(u^2),  
\end{equation}	
thus we need to impose 
\begin{equation}
\partial_u \tilde{\Phi}(u=0,t,x) = C_{\tilde{\Phi}} (t,x) = - \phi_\text{response}(t,x)
\end{equation}	

for $\tilde{\Phi}$ to be in agreement with the asymptotic analysis for $d_+ \Phi$.

\item For $\tilde{B}$, things are a bit trickier. The shift above is the only one that reconciles the expansion of $d_+B$ near the boundary with the value of $g_{\tilde{B}}(t,x)$ obtained from its redefinition; every other choice leads to a contradiction between the equation of motion and its expected behaviour, which requires
\begin{equation}
\tilde{B}(u=0,t,x) = \frac{1}{4} (\partial_x \phi_0(t,x))^2 - \frac{1}{4} (\partial_x f^{(3)}(t,x))^2 \,.
\end{equation}	
\end{itemize}

With those solutions in hand, the next step is to calculate $\partial_t \lambda$. Since $\lambda$ determines the position of the apparent horizon, it makes sense to solve for its dynamics using information about the horizon for increased accuracy. To proceed, we need two equations: the horizon condition (\ref{appcond}), which determines a condition on $d_+ \Sigma$, and the stationarity constraint, which ensures that the horizon condition is satisfied at all times. Rather than using the field redefinition of $d_+ \Sigma$ in the horizon condition, we use the definition of $d_+$ to express $d_+ \Sigma$ as 
\begin{equation}
d_+ \Sigma = \partial_t \lambda + A + d_+\left( -\frac{1}{4} \phi^2\right)
\end{equation}	
As a result of our knowledge of $a_\text{hor}(t,x)$ from the stationarity condition, we can calculate $\partial_t \lambda$, which in turn enables us to solve for $a$ everywhere by using the last relation above.

Now that we have solved for all the fields on a particular time slice, all that is left to do is to propagate $\lambda$, $T_{00}$, $T_{tx}$, $\phi$ and $b$ forward in time (the last two from our knowledge of $d_+ \Phi$ and $d_+ B$), and reiterate the procedure.


\begin{thebibliography}{10}

\bibitem{Calabrese:2005in}
P.~Calabrese and J.~L. Cardy, {\it {Evolution of entanglement entropy in
  one-dimensional systems}},  {\em J.Stat.Mech.} {\bf 0504} (2005) P04010,
  [\href{http://arxiv.org/abs/cond-mat/0503393}{{\tt cond-mat/0503393}}].

\bibitem{Calabrese:2006rx}
P.~Calabrese and J.~L. Cardy, {\it {Time-dependence of correlation functions
  following a quantum quench}},  {\em Phys.Rev.Lett.} {\bf 96} (2006) 136801,
  [\href{http://arxiv.org/abs/cond-mat/0601225}{{\tt cond-mat/0601225}}].

\bibitem{Calabrese:2007rg}
P.~Calabrese and J.~Cardy, {\it {Quantum Quenches in Extended Systems}},  {\em
  J.Stat.Mech.} {\bf 0706} (2007) P06008,
  [\href{http://arxiv.org/abs/0704.1880}{{\tt arXiv:0704.1880}}].

\bibitem{Hubeny:2007xt}
V.~E. Hubeny, M.~Rangamani, and T.~Takayanagi, {\it {A Covariant holographic
  entanglement entropy proposal}},  {\em JHEP} {\bf 0707} (2007) 062,
  [\href{http://arxiv.org/abs/0705.0016}{{\tt arXiv:0705.0016}}].

\bibitem{Chesler:2008hg}
P.~M. Chesler and L.~G. Yaffe, {\it {Horizon formation and far-from-equilibrium
  isotropization in supersymmetric Yang-Mills plasma}},  {\em Phys.Rev.Lett.}
  {\bf 102} (2009) 211601, [\href{http://arxiv.org/abs/0812.2053}{{\tt
  arXiv:0812.2053}}].

\bibitem{Bhattacharyya:2009uu}
S.~Bhattacharyya and S.~Minwalla, {\it {Weak Field Black Hole Formation in
  Asymptotically AdS Spacetimes}},  {\em JHEP} {\bf 0909} (2009) 034,
  [\href{http://arxiv.org/abs/0904.0464}{{\tt arXiv:0904.0464}}].

\bibitem{Das:2010yw}
S.~R. Das, T.~Nishioka, and T.~Takayanagi, {\it {Probe Branes, Time-dependent
  Couplings and Thermalization in AdS/CFT}},  {\em JHEP} {\bf 1007} (2010) 071,
  [\href{http://arxiv.org/abs/1005.3348}{{\tt arXiv:1005.3348}}].

\bibitem{Hubeny:2010ry}
V.~E. Hubeny and M.~Rangamani, {\it {A Holographic view on physics out of
  equilibrium}},  {\em Adv.High Energy Phys.} {\bf 2010} (2010) 297916,
  [\href{http://arxiv.org/abs/1006.3675}{{\tt arXiv:1006.3675}}].

\bibitem{AbajoArrastia:2010yt}
J.~Abajo-Arrastia, J.~Aparicio, and E.~Lopez, {\it {Holographic Evolution of
  Entanglement Entropy}},  {\em JHEP} {\bf 1011} (2010) 149,
  [\href{http://arxiv.org/abs/1006.4090}{{\tt arXiv:1006.4090}}].

\bibitem{Albash:2010mv}
T.~Albash and C.~V. Johnson, {\it {Evolution of Holographic Entanglement
  Entropy after Thermal and Electromagnetic Quenches}},  {\em New J.Phys.} {\bf
  13} (2011) 045017, [\href{http://arxiv.org/abs/1008.3027}{{\tt
  arXiv:1008.3027}}].

\bibitem{Balasubramanian:2010ce}
V.~Balasubramanian, A.~Bernamonti, J.~de~Boer, N.~Copland, B.~Craps, et~al.,
  {\it {Thermalization of Strongly Coupled Field Theories}},  {\em
  Phys.Rev.Lett.} {\bf 106} (2011) 191601,
  [\href{http://arxiv.org/abs/1012.4753}{{\tt arXiv:1012.4753}}].

\bibitem{Balasubramanian:2011ur}
V.~Balasubramanian, A.~Bernamonti, J.~de~Boer, N.~Copland, B.~Craps, et~al.,
  {\it {Holographic Thermalization}},  {\em Phys.Rev.} {\bf D84} (2011) 026010,
  [\href{http://arxiv.org/abs/1103.2683}{{\tt arXiv:1103.2683}}].

\bibitem{Aparicio:2011zy}
J.~Aparicio and E.~Lopez, {\it {Evolution of Two-Point Functions from
  Holography}},  {\em JHEP} {\bf 1112} (2011) 082,
  [\href{http://arxiv.org/abs/1109.3571}{{\tt arXiv:1109.3571}}].

\bibitem{Basu:2011ft}
P.~Basu and S.~R. Das, {\it {Quantum Quench across a Holographic Critical
  Point}},  {\em JHEP} {\bf 1201} (2012) 103,
  [\href{http://arxiv.org/abs/1109.3909}{{\tt arXiv:1109.3909}}].

\bibitem{Balasubramanian:2011at}
V.~Balasubramanian, A.~Bernamonti, N.~Copland, B.~Craps, and F.~Galli, {\it
  {Thermalization of mutual and tripartite information in strongly coupled two
  dimensional conformal field theories}},  {\em Phys.Rev.} {\bf D84} (2011)
  105017, [\href{http://arxiv.org/abs/1110.0488}{{\tt arXiv:1110.0488}}].

\bibitem{Keranen:2011xs}
V.~Keranen, E.~Keski-Vakkuri, and L.~Thorlacius, {\it {Thermalization and
  entanglement following a non-relativistic holographic quench}},  {\em
  Phys.Rev.} {\bf D85} (2012) 026005,
  [\href{http://arxiv.org/abs/1110.5035}{{\tt arXiv:1110.5035}}].

\bibitem{Galante:2012pv}
D.~Galante and M.~Schvellinger, {\it {Thermalization with a chemical potential
  from AdS spaces}},  {\em JHEP} {\bf 1207} (2012) 096,
  [\href{http://arxiv.org/abs/1205.1548}{{\tt arXiv:1205.1548}}].

\bibitem{Caceres:2012em}
E.~Caceres and A.~Kundu, {\it {Holographic Thermalization with Chemical
  Potential}},  {\em JHEP} {\bf 1209} (2012) 055,
  [\href{http://arxiv.org/abs/1205.2354}{{\tt arXiv:1205.2354}}].

\bibitem{Buchel:2012gw}
A.~Buchel, L.~Lehner, and R.~C. Myers, {\it {Thermal quenches in N=2*
  plasmas}},  {\em JHEP} {\bf 1208} (2012) 049,
  [\href{http://arxiv.org/abs/1206.6785}{{\tt arXiv:1206.6785}}].

\bibitem{Bhaseen:2012gg}
M.~Bhaseen, J.~P. Gauntlett, B.~Simons, J.~Sonner, and T.~Wiseman, {\it
  {Holographic Superfluids and the Dynamics of Symmetry Breaking}},  {\em
  Phys.Rev.Lett.} {\bf 110} (2013) 015301,
  [\href{http://arxiv.org/abs/1207.4194}{{\tt arXiv:1207.4194}}].

\bibitem{Basu:2012gg}
P.~Basu, D.~Das, S.~R. Das, and T.~Nishioka, {\it {Quantum Quench Across a Zero
  Temperature Holographic Superfluid Transition}},  {\em JHEP} {\bf 1303}
  (2013) 146, [\href{http://arxiv.org/abs/1211.7076}{{\tt arXiv:1211.7076}}].

\bibitem{Hubeny:2013hz}
V.~E. Hubeny, M.~Rangamani, and E.~Tonni, {\it {Thermalization of Causal
  Holographic Information}},  {\em JHEP} {\bf 1305} (2013) 136,
  [\href{http://arxiv.org/abs/1302.0853}{{\tt arXiv:1302.0853}}].

\bibitem{Nozaki:2013wia}
M.~Nozaki, T.~Numasawa, and T.~Takayanagi, {\it {Holographic Local Quenches and
  Entanglement Density}},  {\em JHEP} {\bf 05} (2013) 080,
  [\href{http://arxiv.org/abs/1302.5703}{{\tt arXiv:1302.5703}}].

\bibitem{Buchel:2013lla}
A.~Buchel, L.~Lehner, R.~C. Myers, and A.~van Niekerk, {\it {Quantum quenches
  of holographic plasmas}},  {\em JHEP} {\bf 1305} (2013) 067,
  [\href{http://arxiv.org/abs/1302.2924}{{\tt arXiv:1302.2924}}].

\bibitem{Hartman:2013qma}
T.~Hartman and J.~Maldacena, {\it {Time Evolution of Entanglement Entropy from
  Black Hole Interiors}},  {\em JHEP} {\bf 1305} (2013) 014,
  [\href{http://arxiv.org/abs/1303.1080}{{\tt arXiv:1303.1080}}].

\bibitem{Basu:2013soa}
P.~Basu, D.~Das, S.~R. Das, and K.~Sengupta, {\it {Quantum Quench and Double
  Trace Couplings}},  {\em JHEP} {\bf 1312} (2013) 070,
  [\href{http://arxiv.org/abs/1308.4061}{{\tt arXiv:1308.4061}}].

\bibitem{Buchel:2013gba}
A.~Buchel, R.~C. Myers, and A.~van Niekerk, {\it {Universality of Abrupt
  Holographic Quenches}},  {\em Phys.Rev.Lett.} {\bf 111} (2013) 201602,
  [\href{http://arxiv.org/abs/1307.4740}{{\tt arXiv:1307.4740}}].

\bibitem{Liu:2013iza}
H.~Liu and S.~J. Suh, {\it {Entanglement Tsunami: Universal Scaling in
  Holographic Thermalization}},  {\em Phys. Rev. Lett.} {\bf 112} (2014)
  011601, [\href{http://arxiv.org/abs/1305.7244}{{\tt arXiv:1305.7244}}].

\bibitem{Balasubramanian:2013oga}
V.~Balasubramanian, A.~Bernamonti, J.~de~Boer, B.~Craps, L.~Franti, et~al.,
  {\it {Inhomogeneous holographic thermalization}},  {\em JHEP} {\bf 1310}
  (2013) 082, [\href{http://arxiv.org/abs/1307.7086}{{\tt arXiv:1307.7086}}].

\bibitem{Liu:2013qca}
H.~Liu and S.~J. Suh, {\it {Entanglement growth during thermalization in
  holographic systems}},  {\em Phys. Rev.} {\bf D89} (2014), no.~6 066012,
  [\href{http://arxiv.org/abs/1311.1200}{{\tt arXiv:1311.1200}}].

\bibitem{Asplund:2013zba}
C.~T. Asplund and A.~Bernamonti, {\it {Mutual information after a local quench
  in conformal field theory}},  {\em Phys. Rev.} {\bf D89} (2014), no.~6
  066015, [\href{http://arxiv.org/abs/1311.4173}{{\tt arXiv:1311.4173}}].

\bibitem{Abajo-Arrastia:2014fma}
J.~Abajo-Arrastia, E.~da~Silva, E.~Lopez, J.~Mas, and A.~Serantes, {\it
  {Holographic Relaxation of Finite Size Isolated Quantum Systems}},  {\em
  JHEP} {\bf 1405} (2014) 126, [\href{http://arxiv.org/abs/1403.2632}{{\tt
  arXiv:1403.2632}}].

\bibitem{Astaneh:2014fga}
A.~F. Astaneh and A.~E. Mosaffa, {\it {Quantum Local Quench, AdS/BCFT and Yo-Yo
  String}},  {\em JHEP} {\bf 05} (2015) 107,
  [\href{http://arxiv.org/abs/1405.5469}{{\tt arXiv:1405.5469}}].

\bibitem{Buchel:2014gta}
A.~Buchel, R.~C. Myers, and A.~van Niekerk, {\it {Nonlocal probes of
  thermalization in holographic quenches with spectral methods}},
  \href{http://arxiv.org/abs/1410.6201}{{\tt arXiv:1410.6201}}.

\bibitem{Das:2014hqa}
S.~R. Das, D.~A. Galante, and R.~C. Myers, {\it {Universality in fast quantum
  quenches}},  {\em JHEP} {\bf 02} (2015) 167,
  [\href{http://arxiv.org/abs/1411.7710}{{\tt arXiv:1411.7710}}].

\bibitem{Bai:2014tla}
X.~Bai, B.-H. Lee, L.~Li, J.-R. Sun, and H.-Q. Zhang, {\it {Time Evolution of
  Entanglement Entropy in Quenched Holographic Superconductors}},  {\em JHEP}
  {\bf 04} (2015) 066, [\href{http://arxiv.org/abs/1412.5500}{{\tt
  arXiv:1412.5500}}].

\bibitem{Rangamani:2015sha}
M.~Rangamani, M.~Rozali, and A.~Wong, {\it {Driven Holographic CFTs}},  {\em
  JHEP} {\bf 04} (2015) 093, [\href{http://arxiv.org/abs/1502.05726}{{\tt
  arXiv:1502.05726}}].

\bibitem{Leichenauer:2015xra}
S.~Leichenauer and M.~Moosa, {\it {Entanglement Tsunami in (1+1)-Dimensions}},
  \href{http://arxiv.org/abs/1505.04225}{{\tt arXiv:1505.04225}}.

\bibitem{Das:2015jka}
S.~R. Das, D.~A. Galante, and R.~C. Myers, {\it {Smooth and fast versus
  instantaneous quenches in quantum field theory}},  {\em JHEP} {\bf 08} (2015)
  073, [\href{http://arxiv.org/abs/1505.05224}{{\tt arXiv:1505.05224}}].

\bibitem{Ecker:2015kna}
C.~Ecker, D.~Grumiller, and S.~A. Stricker, {\it {Evolution of holographic
  entanglement entropy in an anisotropic system}},  {\em JHEP} {\bf 07} (2015)
  146, [\href{http://arxiv.org/abs/1506.02658}{{\tt arXiv:1506.02658}}].

\bibitem{Ziogas:2015aja}
V.~Ziogas, {\it {Holographic mutual information in global Vaidya-BTZ
  spacetime}},  {\em JHEP} {\bf 09} (2015) 114,
  [\href{http://arxiv.org/abs/1507.00306}{{\tt arXiv:1507.00306}}].

\bibitem{Sohrabi:2015qda}
K.~A. Sohrabi, {\it {Inhomogeneous Thermal Quenches}},
  \href{http://arxiv.org/abs/1509.00245}{{\tt arXiv:1509.00245}}.

\bibitem{Camilo:2015wea}
G.~Camilo, B.~Cuadros-Melgar, and E.~Abdalla, {\it {Holographic quenches
  towards a Lifshitz point}},  \href{http://arxiv.org/abs/1511.08843}{{\tt
  arXiv:1511.08843}}.

\bibitem{Ryu:2006bv}
S.~Ryu and T.~Takayanagi, {\it {Holographic derivation of entanglement entropy
  from AdS/CFT}},  {\em Phys.Rev.Lett.} {\bf 96} (2006) 181602,
  [\href{http://arxiv.org/abs/hep-th/0603001}{{\tt hep-th/0603001}}].

\bibitem{Ryu:2006ef}
S.~Ryu and T.~Takayanagi, {\it {Aspects of Holographic Entanglement Entropy}},
  {\em JHEP} {\bf 0608} (2006) 045,
  [\href{http://arxiv.org/abs/hep-th/0605073}{{\tt hep-th/0605073}}].

\bibitem{Danielsson:1999fa}
U.~H. Danielsson, E.~Keski-Vakkuri, and M.~Kruczenski, {\it {Black hole
  formation in AdS and thermalization on the boundary}},  {\em JHEP} {\bf 0002}
  (2000) 039, [\href{http://arxiv.org/abs/hep-th/9912209}{{\tt
  hep-th/9912209}}].

\bibitem{Casini:2015zua}
H.~Casini, H.~Liu, and M.~Mezei, {\it {Spread of entanglement and causality}},
  \href{http://arxiv.org/abs/1509.05044}{{\tt arXiv:1509.05044}}.

\bibitem{Casini:2003ix}
H.~Casini, {\it {Geometric entropy, area, and strong subadditivity}},  {\em
  Class. Quant. Grav.} {\bf 21} (2004) 2351--2378,
  [\href{http://arxiv.org/abs/hep-th/0312238}{{\tt hep-th/0312238}}].

\bibitem{Headrick:2014cta}
M.~Headrick, V.~E. Hubeny, A.~Lawrence, and M.~Rangamani, {\it {Causality \&
  holographic entanglement entropy}},  {\em JHEP} {\bf 1412} (2014) 162,
  [\href{http://arxiv.org/abs/1408.6300}{{\tt arXiv:1408.6300}}].

\bibitem{Asplund:2014coa}
C.~T. Asplund, A.~Bernamonti, F.~Galli, and T.~Hartman, {\it {Holographic
  Entanglement Entropy from 2d CFT: Heavy States and Local Quenches}},  {\em
  JHEP} {\bf 02} (2015) 171, [\href{http://arxiv.org/abs/1410.1392}{{\tt
  arXiv:1410.1392}}].

\bibitem{Asplund:2015eha}
C.~T. Asplund, A.~Bernamonti, F.~Galli, and T.~Hartman, {\it {Entanglement
  Scrambling in 2d Conformal Field Theory}},  {\em JHEP} {\bf 09} (2015) 110,
  [\href{http://arxiv.org/abs/1506.03772}{{\tt arXiv:1506.03772}}].

\bibitem{Bhattacharyya:2008xc} 
  S.~Bhattacharyya, V.~E.~Hubeny, R.~Loganayagam, G.~Mandal, S.~Minwalla, T.~Morita, M.~Rangamani and H.~S.~Reall,
{\it Local Fluid Dynamical Entropy from Gravity},
 {\em JHEP} {\bf 0806} (2008) 055,
[\href{http://arxiv.org/abs/0803.2526}{{\tt arXiv:0803.2526}}].

\bibitem{Hubeny:2013dea}
V.~E. Hubeny and H.~Maxfield, {\it {Holographic probes of collapsing black
  holes}},  {\em JHEP} {\bf 1403} (2014) 097,
  [\href{http://arxiv.org/abs/1312.6887}{{\tt arXiv:1312.6887}}].

\bibitem{Chesler:2013lia}
P.~M. Chesler and L.~G. Yaffe, {\it {Numerical solution of gravitational
  dynamics in asymptotically anti-de Sitter spacetimes}},  {\em JHEP} {\bf
  1407} (2014) 086, [\href{http://arxiv.org/abs/1309.1439}{{\tt
  arXiv:1309.1439}}].

\bibitem{deHaro:2000xn}
S.~de~Haro, S.~N. Solodukhin, and K.~Skenderis, {\it {Holographic
  reconstruction of space-time and renormalization in the AdS / CFT
  correspondence}},  {\em Commun. Math. Phys.} {\bf 217} (2001) 595--622,
  [\href{http://arxiv.org/abs/hep-th/0002230}{{\tt hep-th/0002230}}].

\bibitem{Boyd:2001aa}
J.~P. Boyd, {\em Chebyshev and Fourier spectral methods}.
\newblock Dover Publications, 2001.

\bibitem{Lieb:1972wy}
E.~H. Lieb and D.~W. Robinson, {\it {The finite group velocity of quantum spin
  systems}},  {\em Commun. Math. Phys.} {\bf 28} (1972) 251--257.

\bibitem{Eisler:2007aa}
V.~Eisler and I.~Peschel, {\it {Evolution of entanglement after a local
  quench}},  {\em Journal of Statistical Mechanics: Theory and Experiment} {\bf
  6} (June, 2007) 5, [\href{http://arxiv.org/abs/cond-mat/0703379}{{\tt
  cond-mat/0703379}}].

\bibitem{Hayden:2007cs}
P.~Hayden and J.~Preskill, {\it {Black holes as mirrors: Quantum information in
  random subsystems}},  {\em JHEP} {\bf 09} (2007) 120,
  [\href{http://arxiv.org/abs/0708.4025}{{\tt arXiv:0708.4025}}].

\bibitem{Sekino:2008he}
Y.~Sekino and L.~Susskind, {\it {Fast Scramblers}},  {\em JHEP} {\bf 10} (2008)
  065, [\href{http://arxiv.org/abs/0808.2096}{{\tt arXiv:0808.2096}}].

\bibitem{Cardy:2015xaa}
J.~Cardy, {\it {Quantum Quenches to a Critical Point in One Dimension: some
  further results}},  \href{http://arxiv.org/abs/1507.07266}{{\tt
  arXiv:1507.07266}}.

\bibitem{Rangamani:2016aa}
M.~Rangamani, M.~Rozali, and A.~Vincart-Emard, {\it {work in progress}}.

\bibitem{Balasubramanian:2013yqa}
K.~Balasubramanian and C.~P. Herzog, {\it {Losing Forward Momentum
  Holographically}},  {\em Class.Quant.Grav.} {\bf 31} (2014) 125010,
  [\href{http://arxiv.org/abs/1312.4953}{{\tt arXiv:1312.4953}}].

\end{thebibliography}

\providecommand{\href}[2]{#2}\begingroup\raggedright\endgroup

\end{document}